\documentclass[aps,prc,reprint,unsortedaddress,superscriptaddress,
floatfix,longbibliography,]{revtex4-1}

\usepackage{amsmath,amssymb,amsbsy}
\usepackage{hyperref}
\usepackage[all]{hypcap}
\usepackage{enumerate}
\usepackage{isotope}
\usepackage{lipsum}
\usepackage{bbm}
\usepackage{graphicx}
\usepackage{physics}
\usepackage{feynmp}
\usepackage{multirow,dcolumn}
\newcolumntype{.}{D{.}{.}{1.5}}
\newcolumntype{,}{D{.}{.}{1.8}}
\newcolumntype{;}{D{.}{.}{2.3}}
\newcolumntype{+}{D{.}{.}{5.3}}
\newcolumntype{-}{D{.}{.}{3.4}}

\hypersetup{colorlinks=true,citecolor=[rgb]{0,0,0.8},
linkcolor=[rgb]{0,0,0.8},urlcolor=[rgb]{0,0,0.8}}

\newcommand{\up}{\uparrow}
\newcommand{\down}{\downarrow}

\newcommand{\nn}{\ensuremath{NN}}
\newcommand{\nnn}{\ensuremath{3N}}
\newcommand{\nxlo}[1]{%
   \ifnum0=#1\relax%
      \text{LO}%
   \else%
   \ifnum1=#1\relax%
      \text{NLO}%
   \else%
      \text{N}\ensuremath{^{#1}}\text{LO}%
   \fi
   \fi
}
\newcommand{\nalpha}{$n$-$\alpha$}
\newcommand{\tdott}[2]{\vb*{\tau}_{#1}\vdot\vb*{\tau}_{#2}}
\newcommand{\sdots}[2]{\vb*{\sigma}_{#1}\vdot\vb*{\sigma}_{#2}}
\newcommand{\drtn}[1]{\delta_{R_{\nnn{}}}\!(#1)}
\newcommand{\drzero}[1]{\delta_{R_0}\!(#1)}

\newcommand{\eu}[1]{\mathrm{e}^{#1}}
\newcommand{\ii}{\mathrm{i}}

\begin{document}
\unitlength=1mm
\title{Quantum Monte Carlo calculations of light nuclei with\\ 
local chiral two- and three-nucleon interactions}

\author{J.~E.~Lynn}
\email[E-mail:~]{joel.lynn@gmail.com}
\affiliation{Institut f\"ur Kernphysik,
Technische Universit\"at Darmstadt, 64289 Darmstadt, Germany}
\affiliation{ExtreMe Matter Institute EMMI, GSI Helmholtzzentrum f\"ur
Schwerionenforschung GmbH, 64291 Darmstadt, Germany}
\author{I.~Tews}
\email[E-mail:~]{itews@uw.edu}
\affiliation{Institute for Nuclear Theory,
University of Washington, Seattle, Washington 98195-1550, USA}
\author{J.~Carlson}
\affiliation{Theoretical Division, Los Alamos National Laboratory,
Los Alamos, New Mexico 87545, USA}
\author{S.~Gandolfi}
\affiliation{Theoretical Division, Los Alamos National Laboratory,
Los Alamos, New Mexico 87545, USA}
\author{A.~Gezerlis}
\affiliation{Department of Physics, University of Guelph,
Guelph, Ontario, N1G 2W1, Canada}
\author{K.~E.~Schmidt}
\affiliation{Department of Physics, Arizona State University, Tempe,
Arizona 85287, USA}
\author{A.~Schwenk}
\affiliation{Institut f\"ur Kernphysik,
Technische Universit\"at Darmstadt, 64289 Darmstadt, Germany}
\affiliation{ExtreMe Matter Institute EMMI, GSI Helmholtzzentrum f\"ur
Schwerionenforschung GmbH, 64291 Darmstadt, Germany}
\affiliation{Max-Planck-Institut f\"ur Kernphysik, Saupfercheckweg 1, 
69117 Heidelberg, Germany}

\begin{abstract}
Local chiral effective field theory interactions have recently been
developed and used in the context of quantum Monte Carlo few-
and many-body methods for nuclear physics.
In this work, we go over detailed features of local chiral
nucleon-nucleon interactions and examine their effect on properties of
the deuteron, paying special attention to the perturbativeness of the
expansion.
We then turn to three-nucleon interactions, focusing on operator
ambiguities and their interplay with regulator effects.
We then discuss the nuclear Green's function Monte Carlo method,
going over both wave-function correlations and approximations for the
two- and three-body propagators.
Following this, we present a range of results on light nuclei: Binding
energies and distribution functions are contrasted and compared,
starting from several different microscopic interactions.
\end{abstract}
\maketitle

\section{Introduction}
\label{sec:intro}
Theoretical nuclear physics has undergone a renaissance in recent
decades because of two main developments: The increasing reach
and precision of nuclear many-body methods, and the formulation of 
systematic nuclear interactions based on chiral effective field
theory~(EFT).

\textit{Ab initio} many-body methods in nuclear physics include the
no-core shell model~\cite{barrett2013},
nuclear lattice simulations~\cite{epelbaum2011},
the coupled-cluster method~\cite{hagen2013,hagen2015}, the in-medium
similarity renormalization group
(SRG) method~\cite{hergert2015}, self-consistent Green's function
methods~\cite{carbone2013,soma2014}, and quantum Monte Carlo (QMC)
methods~\cite{carlson2015}.
Among these, QMC methods, which are based on the imaginary-time
evolution of a trial wave function and include the Green's function
Monte Carlo (GFMC) method and the auxiliary-field diffusion Monte Carlo
(AFDMC) method, are notable for their high accuracy across various
physical systems.

In a typical calculation, QMC methods reach uncertainties of~$\sim1\%$.
By design, QMC methods introduce only a limited number of approximations
that can be controlled and accounted for systematically. Both the GFMC 
method and the AFDMC method rely on the diffusion equation 
\begin{equation}
\label{eq:master}
\lim_{\tau\to\infty}\eu{-H\tau}\ket{\Psi_T}\to\ket{\Psi_0}\,,
\end{equation}
where $H$ is the Hamiltonian of the system, $\tau$ is imaginary time,
and $\ket{\Psi_T}$ is a trial state for the system not orthogonal to the
ground state $\ket{\Psi_0}$.
These ``diffusion'' methods solve Eq.~(\ref{eq:master}) stochastically
by casting it as a path integral and sampling the paths using Monte
Carlo methods.
This allows one to extract ground- and low-lying excited-state
properties of nuclear systems with high accuracy.

Furthermore, QMC methods are notable
because they approach the many-body problem with a correlated
wave-function--oriented framework.
For certain nuclear systems, e.g., the Hoyle state of \isotope[12]{C},
many-body methods that rely on basis-set expansions can experience  
difficulties in capturing physics that requires a large number of basis
states to describe, such as clustering effects. For QMC methods,
which rely on a trial wave function to describe the state of interest,
these effects are more straightforward to incorporate.
While the GFMC method has an unfavorable scaling behavior with respect
to the nucleon number $A$, the above-mentioned strengths make QMC
calculations of smaller systems an ideal benchmark for other methods.

Besides the exciting advancements for nuclear many-body methods,
the development of chiral EFT as a tool for the derivation of systematic 
nuclear interactions connected to the underlying theory of strong
interactions, quantum chromodynamics (QCD), represents a major step
forward in nuclear theory. The idea, first presented by Weinberg in the
1990s~\cite{Weinberg1990,Weinberg1991,Weinberg1992}, is to write down
the most general Lagrangian consistent with all the symmetries of the
underlying theory, including the chiral symmetry of low-energy QCD, in 
terms of the relevant degrees of freedom at low energies, i.e., nucleons 
and pions.
Together with a power counting scheme to order the resulting contributions
according to their importance, the result is a low-energy effective
field theory for nuclear forces.
The idea was further developed by van Kolck \textit{et al.} in early
pioneering work~\cite{ordonez1994,vanKolck:1994yi,Ordonez:1995rz}.
The first ``modern'' chiral EFT interactions with a 
$\chi^2/\text{datum}$ around 1 in a fit to \nn{} scattering data were
introduced in the early 2000s by
Entem and Machleidt~\cite{Entem:2003ft} and by
Epelbaum~\textit{et al.}~\cite{EGMN3LO}.

The advantages of the chiral EFT approach to nuclear interactions over
commonly used phenomenological approaches include the ability to 
systematically determine consistent many-body interactions and
electroweak currents, as well as to estimate theoretical uncertainties.
The chiral EFT approach, however, is not without some open problems.
These include, e.g., power counting schemes, residual cutoff
dependences, and associated regulator artifacts.
In the past few years, various groups have investigated several aspects
involved in constructing nuclear forces from chiral EFT, e.g., the
fitting protocol~\cite{Ekstrom:2013,ekstrom2015}, 
regulators~\cite{tews2016,dyhdalo2016}, or uncertainty
estimates~\cite{epelbaum2015}, with the goal of improving predictions
based on chiral interactions.

For many years, chiral EFT interactions could not be implemented 
in QMC methods because these interactions are derived in momentum space 
and are typically nonlocal  while QMC methods rely on local 
interactions. In spite of some work to remedy this
shortcoming~\cite{lynn2012}, it remains technically challenging to develop 
QMC methods that both can use nonlocal interactions and lead to results
without large statistical uncertainties; see also Ref.~\cite{roggero2014} for 
an alternative approach.

In recent years, however, it was realized that all sources of
nonlocality can be removed up to
next-to-next-to-leading order~(\nxlo{2}) in the standard Weinberg power
counting.
This led to the development of local chiral interactions and their
implementation in QMC methods~\cite{Gezerlis:2013ipa,
gezerlis2014,tews2016,piarulli2015,Piar16DeltaNuc} and has allowed for
the first QMC studies of light nuclei, neutron matter, and other light
neutron systems with chiral EFT interactions at \nxlo{2} including
\nnn{} interactions~\cite{lynn2014,lynn2016,Klos:2016fdb,chen2016,
gandolfi2017}.
In this paper, we provide details for the calculations of light nuclei
and present additional results.

The structure of this paper is as follows.
In Sec.~\ref{sec:localcheft}, we discuss how local chiral EFT
interactions have been derived, highlight some interesting features of
these local interactions, and discuss open questions.
In Sec.~\ref{sec:qmc}, we describe the GFMC and AFDMC methods in more
detail and discuss the necessary changes in order to accommodate local
chiral EFT interactions.
In Sec.~\ref{sec:enrgrslts}, we provide a summary of results for light
nuclei obtained with QMC methods and chiral EFT interactions.
Finally, we give a summary in Sec.~\ref{sec:sumandconcl}.

\section{Local Chiral Interactions}
\label{sec:localcheft}

As stated in the introduction, chiral EFT is a systematic way of
organizing nuclear interactions. Based on the most general Lagrangian 
consistent with the symmetries of
QCD, and combined with a power counting scheme, it is possible to
expand nuclear interactions in a series with the expansion parameter
$p/\Lambda_b$, where $p$ is a typical
low-momentum scale in nuclear systems of the order of the pion mass
$m_{\pi}$, and $\Lambda_b \sim 500$ MeV  is the breakdown scale that
determines the range of applicability of the EFT.
Then, nuclear interactions can be arranged as
\begin{align}
V_{\nn{}}=V_{\nn{}}^{(0)}+V_{\nn{}}^{(2)}+V_{\nn{}}^{(3)}+\ldots\,,
\end{align}
where the superscript denotes the chiral order (the power of
$Q\sim p/\Lambda_b$ in the corresponding contributions).
At leading order (\nxlo{0}), $Q^0$, two contributions add to
the nuclear interaction: the one-pion exchange (OPE) and 
momentum-independent short-range contact interactions.
At higher orders, two-pion--exchange interactions (TPE) and
momentum-dependent (derivative) contact interactions appear.
For more details on chiral EFT, see
Refs.~\cite{Epelbaum2009,machleidt2011}.

Because chiral EFT is naturally formulated in momentum space, it can 
contain nonlocal parts by construction. In this section, we review the
strategy to remove all sources of nonlocality, present selected results
for the deuteron, show details of the inclusion of \nnn{} interactions
at \nxlo{2}, and discuss several open questions regarding locality and
regularization.

\subsection{Locality in chiral EFT}
Chiral EFT interactions, with the exception of early
pioneering work~\cite{ordonez1994}, have been developed in momentum
space. We define the incoming (outgoing) single-particle momenta in the
\nn{} sector as $\vb{p}_1$, $\vb{p}_2$ ($\vb{p}_1^\prime$,
$\vb{p}_2^\prime$).
Then the incoming (outgoing) relative momentum $\vb{p}$
($\vb{p}^\prime$), the momentum transfer $\vb{q}$, and momentum transfer
in the exchange channel $\vb{k}$ are defined as
\begin{subequations}
\begin{align}
\vb{p}&\equiv\frac{1}{2}(\vb{p}_1-\vb{p}_2)\,,\ 
\vb{p}^\prime\equiv\frac{1}{2}(\vb{p}_1^\prime-\vb{p}_2^\prime)\,,\\
\vb{q}&\equiv\vb{p}_1-\vb{p}_1^\prime=\vb{p}_2^\prime-\vb{p}_2=
\vb{p}-\vb{p}^\prime\,,\\
\vb{k}&\equiv\frac{1}{2}(\vb{p}+\vb{p}^\prime)\,.
\end{align}
\end{subequations}
The Fourier transformation of a function of $\vb{q}$ leads to a local
function in coordinate space that depends only on the two-particle
distance $\vb{r}$, whereas a function of $\vb{k}$ does not.

Chiral EFT \nn{} interactions depend on two linearly independent momenta
out of the four possible momenta ($\vb{p}$ and $\vb{p}'$ or $\vb{q}$
and $\vb{k}$).
There are two possible sources of nonlocality ($\vb{k}$ dependence):
\begin{enumerate}
\item
The momentum-space regulator functions used to regulate high-momentum 
contributions to the interaction and
\item
momentum-dependent higher order contact operators.
\end{enumerate}
We review the method to remove these sources of nonlocality, which 
was first discussed in Ref.~\cite{freunek2007} and later employed in
practice in Refs.~\cite{Gezerlis:2013ipa, gezerlis2014}. 
\subsubsection{Local regulators}
When employing chiral EFT interactions in few- and many-body
calculations, momentum-dependent regulator functions need to be
introduced to cutoff divergences from high-momentum modes.
The typical functional form employed to regulate both the short-range
contact interactions and long-range pion exchanges in nonlocally
regulated chiral EFT interactions is
\begin{equation}
\label{eq:momreg}
f(p^2)=\exp\left[-(p^2/\Lambda_{\nn{}}^2)^n\right]\,,
\end{equation}
where $\Lambda_{\nn{}}$ is the momentum-space cutoff for the \nn{}
sector of the interaction and $n$ is an integer.
Then, the interaction $V(\vb{p},\vb{p}^\prime)$ is regulated as
\begin{equation}
V(\vb{p},\vb{p}^\prime)\to
V(\vb{p},\vb{p}^\prime)f(p^2)f(p^{\prime\,2})\,.
\end{equation}
Even when these regulators are applied to a local interaction
$V(\vb{p},\vb{p}^\prime)=V(\vb{q})$, e.g., a momentum-independent
contact interaction or the local one-pion-exchange interaction, the 
regularized interaction becomes nonlocal due to the explicit $\vb{k}$ 
dependence of the regulator functions.

\begin{figure}[t]
   \includegraphics[width=\columnwidth]{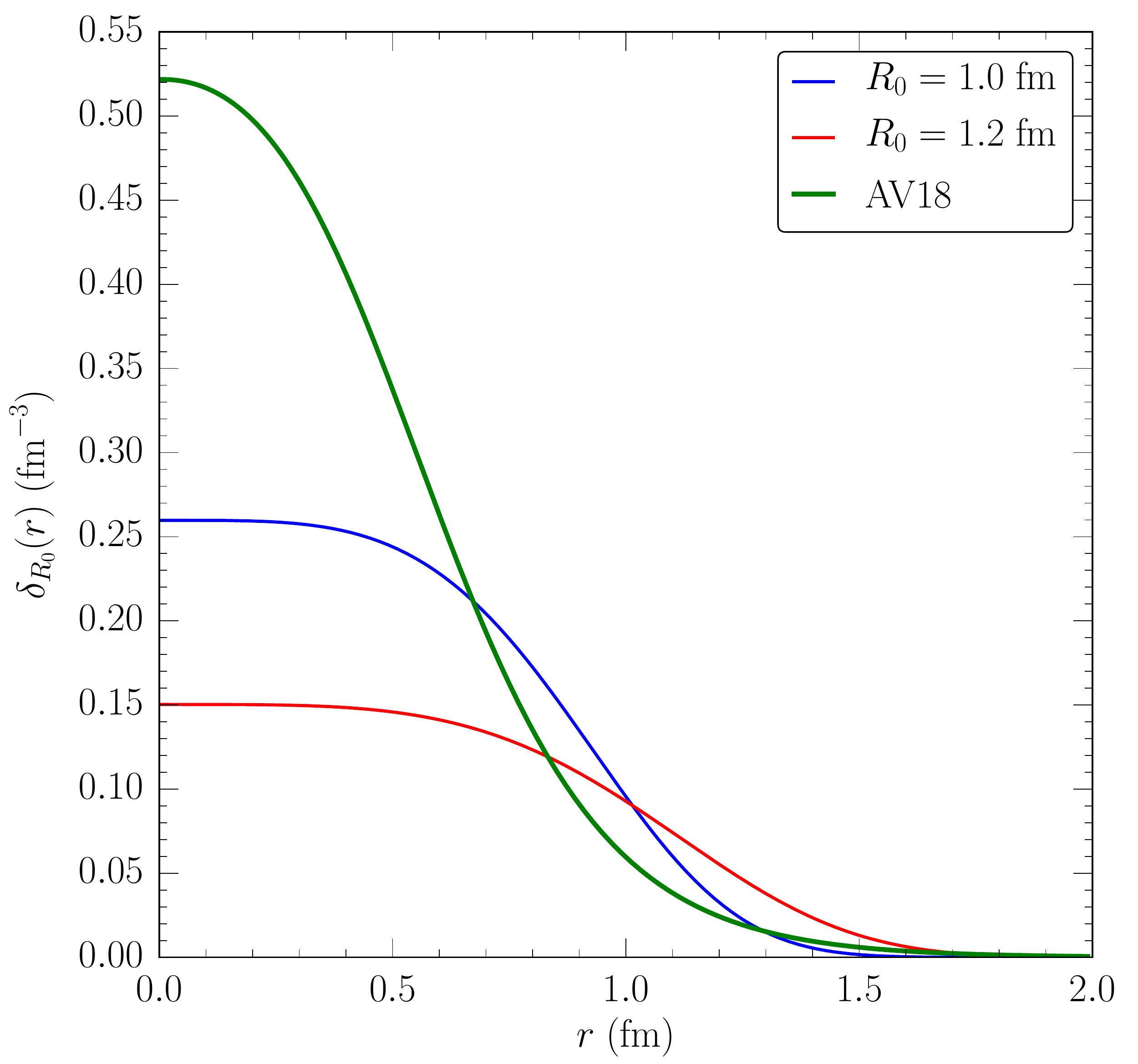}
   \caption{\label{fig:shrtrngreg}
   The (normalized) regulator functions for the short-range
   contact contributions to the local chiral interactions with the
   typical low (hard, $R_0=1.0$ fm) and high (soft, $R_0=1.2$ fm)
   coordinate-space cutoffs.
   In addition, we show the Woods-Saxon core for the central part of the
   Argonne~$v_{18}$ interaction for deuteron pairs.
   (See text for details.)
   }
\end{figure}

A possible solution is to introduce local short- and long-range
regulators. In our case, we regulate the chiral interactions directly 
in coordinate space.
Short-range contact interactions, which Fourier transform to
$\delta$~functions in coordinate space, are regulated by ``smearing them
out,'' i.e., 
\begin{equation}
\label{eq:smeareddelta}
\delta^{(3)}(\vb{r})\to\drzero{r}=
\frac{\eu{-(r/R_0)^n}}{\tfrac{4\pi}{n}\Gamma(\tfrac{3}{n})R_0^3}\,.
\end{equation}
In this work, we choose $n=4$.
The constant $R_0$ serves as a coordinate-space cutoff parameter.
The normalization is chosen such that
\begin{equation}
\label{eq:norm}
\int\dd[3]{r}\drzero{r}=1\,.
\end{equation}
For the long-range parts of the interaction, we use a similar functional
form:
\begin{equation}
\label{eq:longrangereg}
f_\text{long}(r)=1-\eu{-(r/R_0)^4}\,.
\end{equation}
In Fig.~\ref{fig:shrtrngreg}, we compare the short-range regulator used
in the local chiral interactions for two values of the cutoff parameter
$R_0$ with the short-range part used in the Argonne $v_{18}$
interaction~\cite{wiringa1995}.
Specifically, the short-range part of the Argonne $v_{18}$ interaction
is given by $[P^i_{ST,{\nn{}}}+\mu rQ^i_{ST,{\nn{}}}+
(\mu r)^2R^i_{ST,{\nn{}}}]W(r)$, where $\mu$ is the average pion mass;
$P$, $Q$, and $R$ are a set of parameters; and $W(r)$ is a Woods-Saxon
potential.
We display this short-range part of the Argonne $v_{18}$ interaction in
the central channel for deuteron-like pairs, $i=c$ (central), $ST=01$,
and $\nn{}=np$, and normalize as in Eq.~(\ref{eq:norm}); see
Ref.~\cite{wiringa1995} for details on the values of the parameters
$P$, $Q$, $R$, and $\mu$ and the Woods-Saxon potential $W(r)$.

Regarding the range of cutoff parameters, one would like to take 
$R_0$ as small as possible in coordinate space to minimize regulator
artifacts.
However, as has been argued in Ref.~\cite{baru2012} in the context of
the multiple-scattering series, the chiral expansion for the 
pion-exchange potentials breaks down for distances of $r\sim0.8$~fm. 
For $r\gtrsim1.0$~fm, the convergence of the multiple-scattering series, 
however, is  found
to be rather fast. Taking $R_0$ to be arbitrarily large, on the other
hand, cuts off long-range pion physics that is resolved.
We therefore adopt the range 1.0--1.2~fm for the cutoff $R_0$.

Although we stress that there is no direct correspondence between 
coordinate- and momentum-space cutoffs, a possibility of comparing 
the coordinate-space cutoff $R_0$ with typical momentum-space 
cutoff parameters $\Lambda_{\nn{}}$ can be obtained by Fourier
transforming the coordinate-space regulator 
function Eq.~(\ref{eq:smeareddelta}), integrating over all momenta, and
identifying the result with a sharp cutoff.
This gives $\Lambda_{\nn{}}=\hbar c[6\pi^2\drzero{0}]^{1/3}$,
and thus we identify the corresponding momentum scales $\sim500$~MeV
with $R_0=1.0$~fm, and $\sim400$~MeV with $R_0=1.2$~fm.
While a clear translation between coordinate-space and momentum-space
cutoffs can only be obtained when looking at a particular system
or channel, we
note that the estimated range encompassed by our cutoff choice is
typical of other nonlocal chiral EFT interactions; see
also~\cite{Hoppe:2017lok}

Regarding the long-range regulator, there are additional advantages in
choosing a local regulator function. As has been argued  
recently~\cite{epelbaum2015}, the standard
regulator choice Eq.~(\ref{eq:momreg}) distorts the analytic structure of the
partial-wave amplitude near threshold.
Since the long-range interactions in chiral EFT are local [with the
exception of relativistic corrections entering at
next-to-next-to-next-to-leading order (\nxlo{3})], it is logical to
employ a local regulator in coordinate space, which cuts off the
short-range part of the pion-exchange interactions but leaves the
long-range part undisturbed.
For this reason, a (different) local long-range regulator function is
also chosen in the semilocal interactions of
Epelbaum~\textit{et al.}~\cite{epelbaum2015,epelbaum2015prl}.

To regularize pion loops in the TPE contributions at \nxlo{1} and higher
orders, we use the framework of spectral function regularization (SFR). 
In SFR, the integrals over loop momenta in the spectral 
representation of the TPE contributions are cut off at $\tilde{\Lambda}$.
In the following, we use the SFR cutoff $\tilde{\Lambda}=1000$~MeV since 
only a negligible dependence on its choice was 
found~\cite{gezerlis2014,lynn2014}. In particular, increasing 
the SFR cutoff from 1 to 1.4~GeV lowered the \isotope[4]{He} binding 
energy and the energy per particle of pure neutron matter (with only
\nn{} interactions in both cases) by less than $\sim2\%$, which is well
within the $\sim5\%$ truncation uncertainty at this order.
\subsubsection{Local contact operators}
Choosing local regulators removes the first source of nonlocality in
chiral interactions.
The second source of nonlocality originates in the momentum dependence
of higher order contact interactions. Up to \nxlo{2}, these can be
eliminated by exploiting Fierz ambiguities.
At next-to-leading order (\nxlo{1}), i.e., $Q^2$ in the chiral
expansion, the general set of contact operators consistent with all the
symmetries contains 14 different operators.
In addition to spin-isospin dependences, these operators contain
momentum dependences of the form $q^2$ and $k^2$ or
$\vb{q}\times\vb{k}$, where the $k^2$ dependences are undesirable for
local interactions.
One can show using the Pauli principle that between antisymmetric states 
only 7 out of the 14 operators are linearly independent. 
Six linearly independent operators can be chosen to 
be local ($q^2$ dependent) while the 7th operator can be chosen to be
the spin-orbit interaction; see Ref.~\cite{gezerlis2014} for more details.

At \nxlo{3}, there are an additional 15 linearly independent contact
operators.
Only 8 of these are local, while the other 7 operators contain $\vb{k}$
dependences that cannot be removed. 
Nevertheless, it is possible to construct maximally local \nxlo{3}
interactions that contain, at most, nonlocalities of second order in
momentum; see Ref.~\cite{piarulli2015} for initial work in this direction.
To summarize, by choosing an appropriate set of contact operators
and local regulator functions, all sources of nonlocality in chiral EFT can 
be removed up to \nxlo{2}. 
\subsubsection{Uncertainty estimates}
To estimate the truncation uncertainty of the chiral expansion, we
follow Ref.~\cite{epelbaum2015} and estimate the uncertainty of an
observable $X$ at \nxlo{2} as 
\begin{equation}
\begin{split}
\Delta X^{\nxlo{2}}=\max&
\left(\vphantom{X^{\nxlo{2}}}Q^{4} \left|X^{\nxlo{0}}\right|,Q^2
\left|X^{\nxlo{1}}-X^{\nxlo{0}}\right|,\right.\\
&\left.Q\left|X^{\nxlo{2}}-X^{\nxlo{1}}\right|
\right)\,,
\end{split}
\end{equation}
and correspondingly at lower orders.
Furthermore, we require the uncertainties to be at least the size of
the actual higher order corrections. We define the scale $Q$ as $Q=\max
(p/\Lambda_b,m_{\pi}/\Lambda_b)$ with $p$ being a typical momentum
scale of the system. For the work we present below, for nuclei, we 
choose $Q=m_\pi/\Lambda_b$, whereas for our neutron 
matter results, we take $Q$ from the average momentum in a Fermi 
gas $Q=\sqrt{3/5}k_F/\Lambda_b$, with Fermi momentum
$k_F$; see Ref.~\cite{lynn2016}. This choice is
conservative, because typical binding momenta in nuclei are smaller than
the pion mass. These uncertainty estimates provide a quantitative 
estimate of the effect of truncating the chiral expansion at some order
$\nu$.
A careful statistical analysis using Bayesian procedures has been
undertaken in Ref.~\cite{furnstahl2015}, where it was shown that the
prescription we use, first introduced in Ref.~\cite{epelbaum2015},
results in $\nu/(\nu+1)\times100\%$ degree-of-belief (DOB) intervals.
That is, our \nxlo{1} and \nxlo{2} uncertainty estimates are
equivalent to $50\%$ and $\sim67\%$ DOB intervals.

Further details of the \nn{} interaction, e.g., on the inclusion of 
charge-independence and charge-symmetry breaking terms, 
the values of the fitted low-energy constants~(LECs), and phase 
shifts, are given in Ref.~\cite{gezerlis2014}.

\subsection{Deuteron properties}
The deuteron is the lightest nucleus with $A>1$ in nature and provides 
a natural testing ground for the \nn{} interaction.
In this section, we present some properties of this simple
system using chiral interactions at \nxlo{2}. The deuteron wave function 
can be written in terms of its $S$- [$u(r)$]
and $D$-wave [$w(r)$] components as
\begin{equation}
\psi_d^{(M_J)}(\vb{r})=\left[\frac{u(r)}{r}+
\frac{S_{12}(\vb{\hat{r}})}{\sqrt{8}}\frac{w(r)}{r}\right]
\frac{\chi_{M_J}}{\sqrt{4\pi}}\,,
\end{equation}
where $\chi_{M_J}$ is the spin wave function for the total angular 
momentum projection $M_J$, and $S_{ik}(\vb{r})
=3\vb*{\sigma}_i\vdot\vb{\hat{r}}\vb*{\sigma}_k\vdot\vb{\hat{r}}
-\sdots{i}{k}$ is the tensor operator.
The $S$- and $D$-wave components are normalized such that
\begin{equation}
\int_0^\infty
\dd rr^2\left[\left(\frac{u(r)}{r}\right)^2
+\left(\frac{w(r)}{r}\right)^2\right]=1\,.
\end{equation}
The $S$- and $D$-wave components in momentum space are then related 
by Fourier-Bessel transforms
\begin{subequations}
\begin{align}
\frac{\tilde{u}(q)}{q}&=4\pi\int_0^\infty\dd rr^2j_0(qr)\frac{u(r)}{r}\,,\\
\frac{\tilde{w}(q)}{q}&=4\pi\int_0^\infty\dd rr^2j_2(qr)\frac{w(r)}{r}\,,
\end{align}
\end{subequations}
(where $j_l(x)$ is a spherical Bessel function) so that the
normalization is
\begin{equation}
\int_0^\infty\frac{\dd qq^2}{(2\pi)^3}
\left[\left(\frac{\tilde{u}(q)}{q}\right)^2
+\left(\frac{\tilde{w}(q)}{q}\right)^2\right]=1\,.
\end{equation}
\begin{figure}[t]
   \includegraphics[width=\columnwidth]{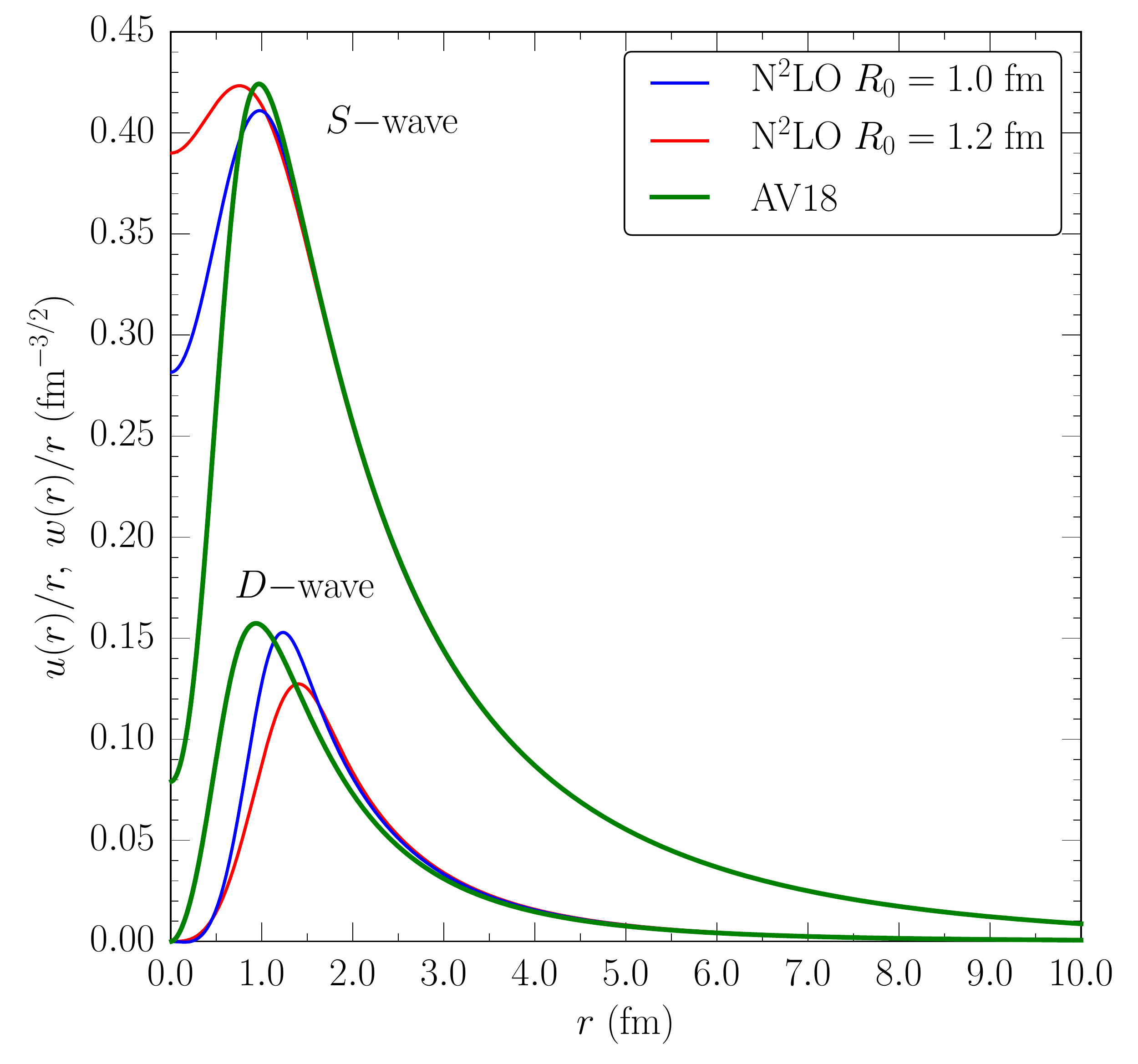}
   \caption{\label{fig:deutwvfunc}
   The deuteron wave functions with $L=0$ ($S$-wave) and $L=2$
   ($D$-wave) at \nxlo{2} for $R_0=1.0$~fm and $R_0=1.2$~fm.
   Also shown are the deuteron wave functions for the Argonne $v_{18}$
   interaction.
   }
\end{figure}

We show the $S$- and $D$-wave components of the deuteron wave 
function in Fig.~\ref{fig:deutwvfunc} for chiral interactions at \nxlo{2} 
with two different cutoff scales along with the deuteron wave function
 for the Argonne $v_{18}$ interaction.
Compared to the hard Argonne $v_{18}$ interaction, the $S$-wave
components of the local chiral interactions are softer, reflected in the
larger value at vanishing pair separation $r$.
As a result, the $D$-wave component is pushed away from $r=0$.
In addition, the $D$-wave component at \nxlo{2} with cutoff $R_0=1.0$~fm
($R_0=1.2$~fm) has a node at $\sim0.2$~fm ($\sim0.02$~fm).
This node has no physical consequences for the deuteron structure and
for both cutoffs occurs at very short distances, where the uncertainty
coming from the truncation of the chiral expansion is largest.

In Table~\ref{tab:deutprops}, we collect a number of properties of the
deuteron at \nxlo{2} and compare with experiment. The deuteron
binding energy is not used in fits of the LECs and can be used as a
check for the local potentials. At \nxlo{2}, the deuteron binding energy
is consistent with experiment taking into account the uncertainties.

\begin{table}
\caption{\label{tab:deutprops}Deuteron properties including the binding
energy $E_b$, asymptotic $D/S$ ratio $\eta_d$, quadrupole moment $Q_d$
(impulse approximation), and root-mean-square (rms) matter radius
$\sqrt{\ev{r^2_d}}$.
Electromagnetic interaction effects are neglected here (when included they change
the values below only within the uncertainties).
The uncertainties for the local chiral interactions represent the
discussed truncation error estimate. See text for more details.
Experimental values are from Refs.~\cite{vanderleun1982,rodning1990,
bishop1979,simon1981}.}
{\renewcommand{\arraystretch}{1.30}
\begin{ruledtabular}
\begin{tabular}{l..,}
&\multicolumn{1}{c}{$R_0=1.0$~fm}&\multicolumn{1}{c}{$R_0=1.2$~fm}&
\multicolumn{1}{c}{Exp}\\
\hline
$E_b$ (MeV)&2.21(2)&2.20(3)&2.224575(9)\\
$\eta_d$&0.0263(3)&0.0267(6)&0.0256(4)\\
$Q_d\ (\text{fm}^2)$&0.286(5)&0.289(6)&0.2859(3)\\
$\sqrt{\ev{r^2_d}}$ (fm)&1.97(2)&1.97(3)&1.9660(68)\\
\end{tabular}
\end{ruledtabular}}
\end{table}

\subsubsection{Momentum distribution}
\label{subsubsec:momdist}
\begin{figure}[b]
   \includegraphics[width=\columnwidth]{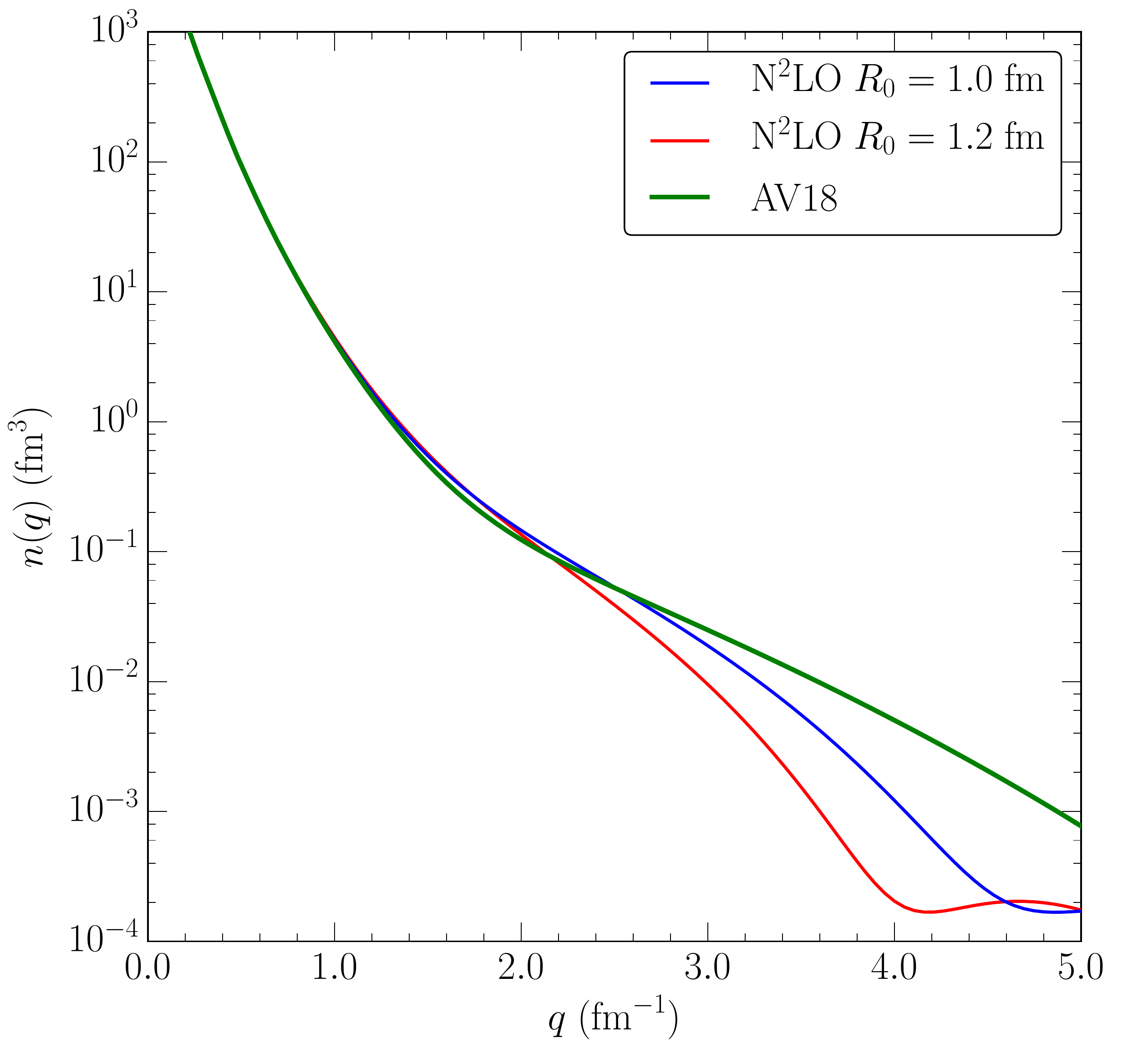}
   \caption{\label{fig:deutmom}
   The deuteron momentum distributions at \nxlo{2} for the two different 
   cutoff scales we use.
   Also shown is the deuteron momentum distribution for the Argonne
   $v_{18}$ interaction.
   }
\end{figure}

The deuteron momentum distribution can be written in terms of the $S$-
and $D$-wave components as
\begin{equation}
n(q)=\frac{1}{4\pi}\left[\left(\frac{\tilde{u}(q)}{q}\right)^2
+\left(\frac{\tilde{w}(q)}{q}\right)^2\right]\,,
\end{equation}
so that the normalization is
\begin{equation}
\int\frac{\dd[3]{q}}{(2\pi)^3}n(q)=1\,.
\end{equation}

In Fig.~\ref{fig:deutmom}, we show the deuteron momentum distribution
for our two cutoff choices along with the momentum distribution obtained
for the Argonne $v_{18}$ interaction.
It is interesting to note that the three momentum distributions display
very similar behavior up to the respective cutoffs of the two chiral
interactions. For
$R_0=1.0\text{ fm}\sim500\text{ MeV}\approx2.5\text{ fm}^{-1}$, the blue
curve begins to deviate significantly from the Argonne $v_{18}$ result
at momenta $\sim 2.5\text{ fm}^{-1}$, while for
$R_0=1.2\text{ fm}\sim400\text{ MeV}\approx2.0\text{ fm}^{-1}$, the red
curve begins to deviate significantly from the Argonne $v_{18}$ result at
$\sim 2.0\text{ fm}^{-1}$.
However, we also emphasize that momentum distributions are necessarily
renormalization scale and scheme dependent and are thus not
observable~\cite{furnstahl2002}.

\subsubsection{Tensor polarization}
\label{subsubsec:deuttens}
Since momentum distributions are scheme and scale dependent, we now
consider the tensor polarization.
The charge form factors for different $M_J$ states are given by
\begin{equation}
F_{C,M_J}(q)=\frac{1}{2}\int\dd[3]{r^\prime}
\rho_{d}^{(M_J)}(\vb{r}^\prime)\eu{\ii{}\vb{q}\vdot\vb{r}^\prime}\,,
\end{equation}
with the deuteron two-body density $\rho_d^{(M_J)}(\vb{r}^\prime)$ in
state $M_J$ in terms of the distance $\vb{r}^\prime$ from the center of
mass:
\begin{figure}[t]
   \includegraphics[width=\columnwidth]{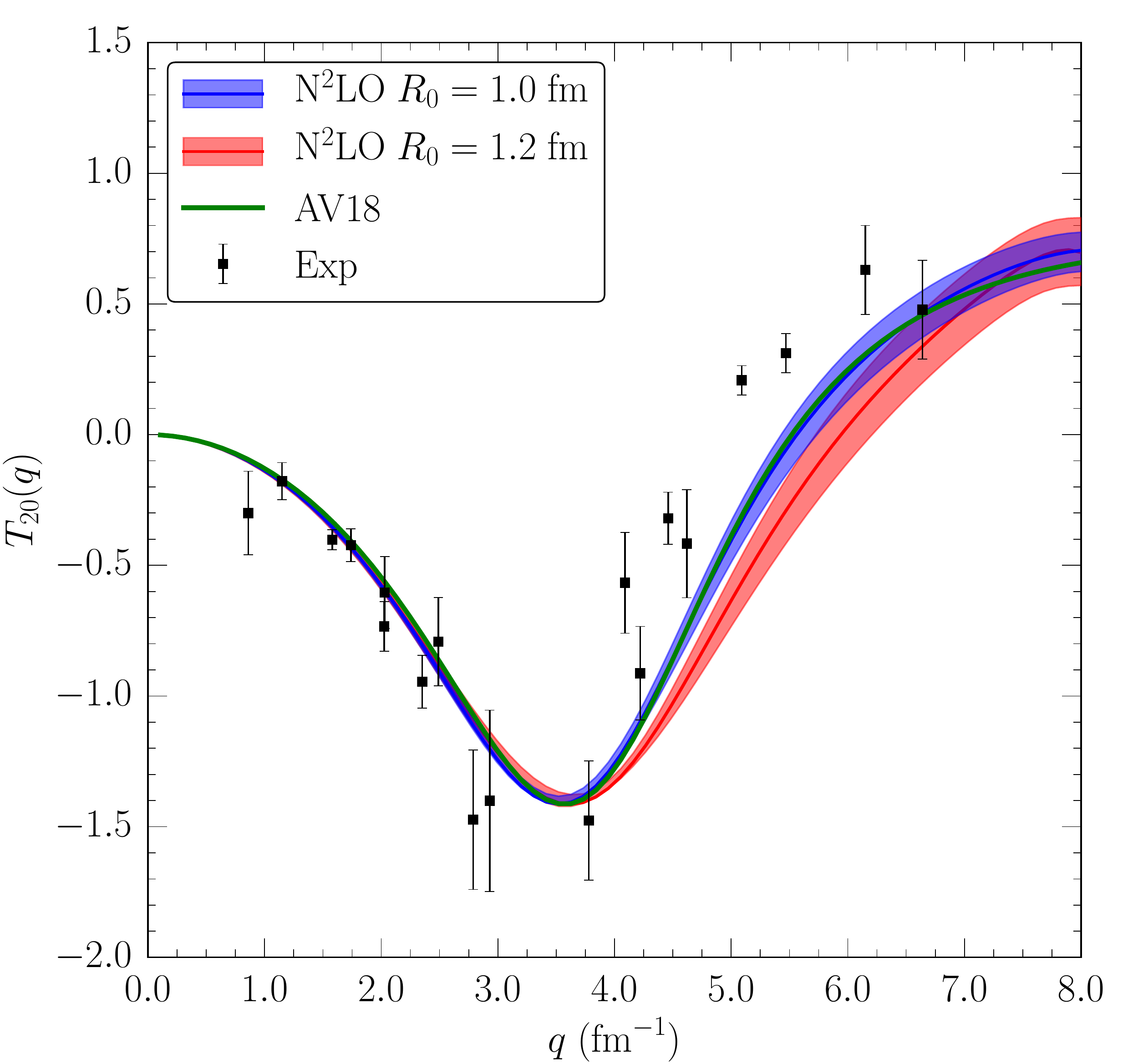}
   \caption{\label{fig:deuttenspoldat}
   The deuteron tensor polarization at \nxlo{2} for the two different 
   cutoff scales we use.
   The bands correspond to an estimate for the uncertainty coming
   from the
   truncation of the chiral expansion as described in the text.
   Also shown is the deuteron tensor polarization for the Argonne
   $v_{18}$ interaction.
   The experimental data are from Ref.~\cite{abbott2000}.
   }
\end{figure}

\begin{subequations}
\begin{align}
\rho_d^{(0)}(\vb{r}^\prime)=&\frac{4}{\pi}
\left[C_0(2r^\prime)-2C_2(2r^\prime)P_2(\cos\theta)\right]\,,\\
\rho_d^{(\pm1)}(\vb{r}^\prime)=&\frac{4}{\pi}
\left[C_0(2r^\prime)+C_2(2r^\prime)P_2(\cos\theta)\right]\,.
\end{align}
\end{subequations}
The functions $C_0$ and $C_2$ are in turn written in terms of the $S$-
and $D$-wave components of the deuteron wave function:
\begin{subequations}
\begin{align}
C_0(r)&=\left(\frac{u(r)}{r}\right)^2+\left(\frac{w(r)}{r}\right)^2\,,\\
C_2(r)&=\sqrt{2}\left(\frac{u(r)}{r}\right)\left(\frac{w(r)}{r}\right)-
\frac{1}{2}+\left(\frac{w(r)}{r}\right)^2\,.
\end{align}
\end{subequations}
The tensor polarization $T_{20}(q)$ is defined (in the impulse
approximation) by~\cite{forest1996}
\begin{equation}
T_{20}(q)\approx-\sqrt{2}
\frac{F_{C,0}^2(q)-F_{C,1}^2(q)}{F_{C,0}^2(q)+2F_{C,1}^2(q)}\,.
\end{equation}
We compare the tensor polarization for both cutoffs and for the Argonne
$v_{18}$ interaction with experimental data~\cite{abbott2000} in
Fig.~\ref{fig:deuttenspoldat}.
The first minimum of $T_{20}(q)$ is experimentally known at
$q~\approx~3.5(5)\text{ fm}^{-1}$~\cite{forest1996,abbott2000,
blast2011}, in agreement with the predictions of all three cases
displayed.
At higher values of $q$, we expect some disagreement between our
calculations and experiment given that we work in the impulse
approximation.
\begin{figure}[t]
   \includegraphics[width=\columnwidth]{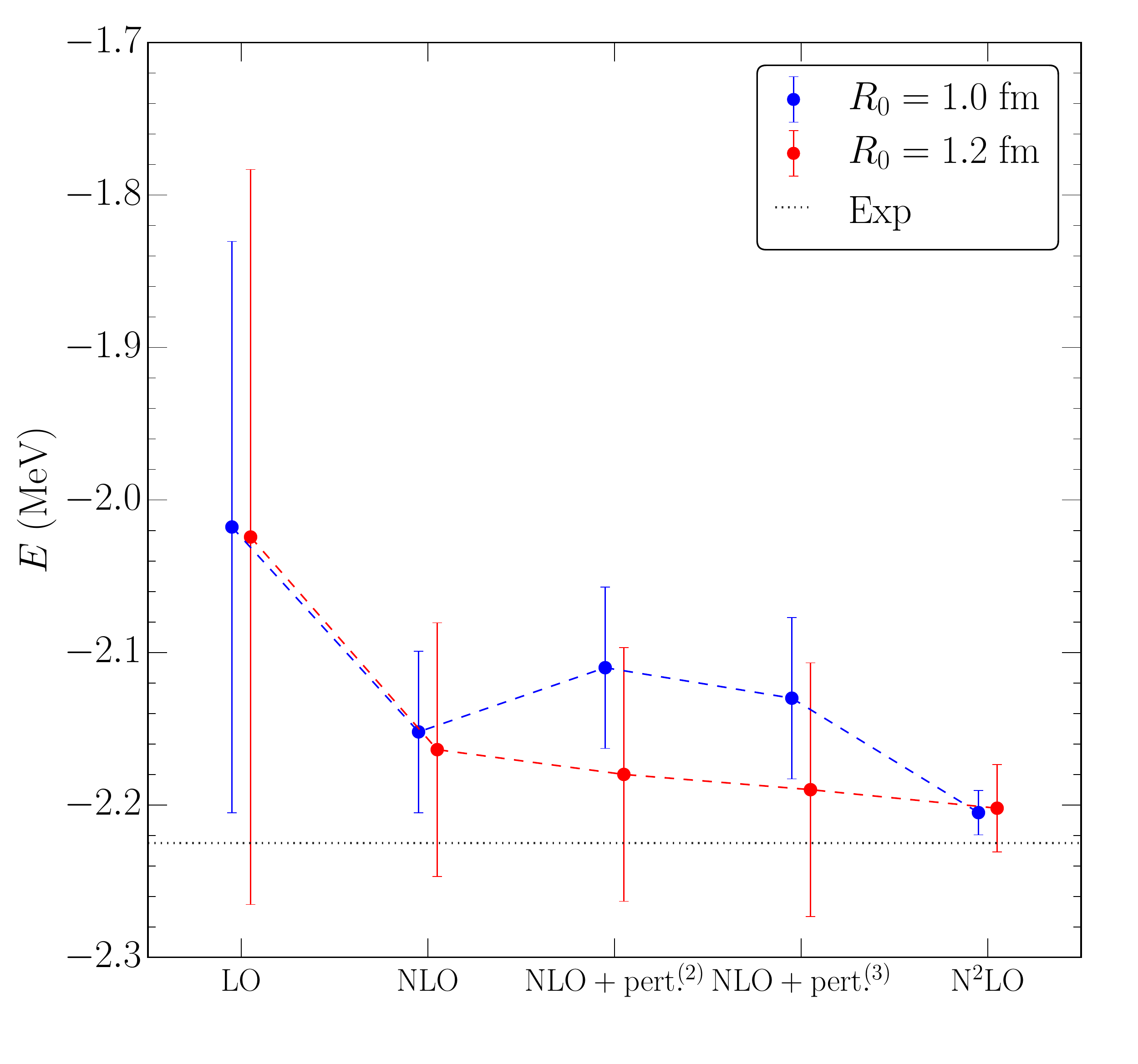}
   \caption{\label{fig:deutpert}
   The deuteron energy at \nxlo{0}, \nxlo{1}, and \nxlo{2} for
   $R_0=1.0$~fm (1.2~fm) in blue (red).
   The error bars are the uncertainty estimates coming from the
   truncation of the chiral expansion as described in the text.
   Also shown, between the \nxlo{1} and \nxlo{2} results, are second- 
   and third-order perturbation theory calculations for the \nxlo{2}
   deuteron energies, taking $H_{\nxlo{1}}$ as the unperturbed
   Hamiltonian, and treating $V_{\nxlo{2}}-V_{\nxlo{1}}$ as a
   perturbation.
   For the perturbation-theory calculations, we take as the uncertainty
   the same estimate as for the \nxlo{1} calculations.
   The dashed lines serve as guides to the eye.
   The horizontal dotted line is the experimental binding energy.
   }
\end{figure}

\subsubsection{Perturbation-theory calculations}
\label{subsubsec:deutpert}
The chiral expansion is meant to be a perturbative expansion in powers
of a small parameter $Q\sim p/\Lambda_b$.
One may well ask if the expected perturbative expansion is evident in
the interactions themselves.
To investigate this, we treat the difference of the
\nxlo{2} and the \nxlo{1} interactions as a perturbation
\begin{equation}
V_\text{pert}\equiv V_{\nxlo{2}}-V_{\nxlo{1}}
\end{equation}
and perform first-, second-, and third-order perturbation-theory
calculations for the deuteron binding energy.
For example, at first order,
\begin{multline}
\ev{H_{\nxlo{1}}+V_\text{pert}}{\psi_d^{({\nxlo{1}})}}\\
=E_{\nxlo{1}}+\ev{V_\text{pert}}{\psi_d^{({\nxlo{1}})}}\,.
\end{multline}

These results at second order and above are displayed
in Fig.~\ref{fig:deutpert}.
As is evident from the figure, the softer interaction with $R_0=1.2$~fm
is more perturbative than the harder interaction with $R_0=1.0$~fm.
In both cases, the perturbative series appears to be converging to the
value at \nxlo{2}, but the convergence is faster for the $R_0=1.2$~fm cutoff.

\subsection{Three-nucleon interactions at
\texorpdfstring{N$^{\boldsymbol{2}}$LO}{N2LO}}
\label{subsec:nnn}
Phenomenological models for \nnn{} interactions, including the 
Urbana~\cite{pudliner1997}, Illinois~\cite{pieper2001}, and
Tucson-Melbourne~\cite{coon2001} models, have been very successfully
used in QMC calculations of nuclear systems. These models are based
on the \nnn{} TPE interaction that was first proposed
by Fujita and Miyazawa nearly 60 years ago~\cite{fujita1957}.
Despite their undeniable success, they suffer from several
shortcomings: They do not emerge naturally from the phenomenological
\nn{} interactions and they are not systematically improvable.

\begin{figure}[t]
   \includegraphics{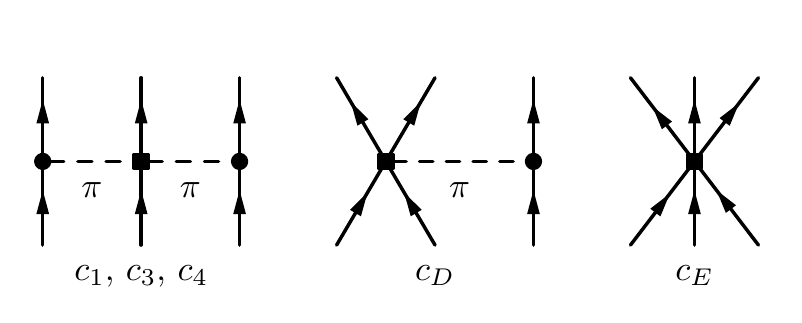}
   \caption{\label{fig:feyndiag}
   The diagrams contributing to \nnn{} interactions
   at \nxlo{2}. Solid lines are nucleons; dashed lines are pions.}
\end{figure}

In chiral EFT, however, \nnn{} interactions naturally emerge in the
expansion and are consistent with the \nn{} interactions. 
Furthermore, they are systematically improvable.
The leading \nnn{} interactions appear at \nxlo{2} in Weinberg
power counting and can be visualized in terms of the diagrams 
in Fig.~\ref{fig:feyndiag}.
The first diagram, proportional to the pion-nucleon LECs $c_1$, $c_3$,
and $c_4$, corresponds to the long-range $S$- and $P$-wave TPE 
interactions by Fujita and Miyazawa.
The LECs $c_i$ already appear in the subleading TPE
interactions at the \nn{} level at the same chiral order, which 
highlights the consistency of the \nn{} and \nnn{} interactions in
chiral EFT.
The second diagram, proportional to the LEC $c_D$, is an
intermediate-range one-pion-exchange--contact interaction, and
the third diagram, proportional to the LEC $c_E$, is a \nnn{}
contact interaction.

The diagrams in Fig.~\ref{fig:feyndiag} give rise to the following
momentum-space \nnn{} interactions:
\begin{subequations}
\begin{align}
\label{subeq:qf1}
V_C&=\frac{1}{2}\left(\frac{g_A}{2F_\pi}\right)^2
\sum_{\pi(ijk)}\frac{(\vb*{\sigma}_i\vdot\vb{q}_i)
(\vb*{\sigma}_j\vdot\vb{q}_j)}{(\vb{q}_i^2+m_\pi^2)
(\vb{q}_j^2+m_\pi^2)}F_{ijk}^{\alpha\beta}\tau_i^\alpha\tau_j^\beta\,,\\
\label{subeq:qf2}
V_D&=-\frac{g_A}{8F_\pi^2}\frac{c_D}{F_\pi^2\Lambda_\chi}
\sum_{\pi(ijk)}\frac{\vb*{\sigma}_j\vdot\vb{q}_j}{\vb{q}_j^2+m_\pi^2}
(\tdott{i}{j})(\vb*{\sigma}_i\vdot\vb{q}_j)\,,\\
\label{subeq:qf3}
V_E&=\frac{c_E}{2F_\pi^4\Lambda_\chi}\sum_{i\ne j}
\tdott{i}{j}\,,
\end{align}
\end{subequations}
\begin{figure}[h]
   \includegraphics{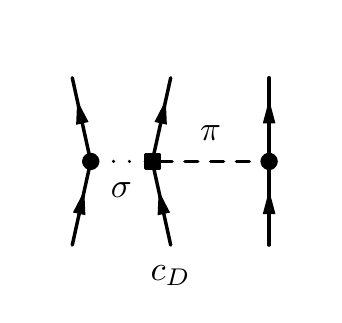}
   \caption{\label{fig:feyndiagftcd2}
   The $c_D$-dependent diagram with a fictitious heavy scalar
   particle $\sigma$ exchanged between two of the nucleons making 
   the participants in the pion exchange explicit.
   Solid lines are nucleons, the dashed line is a pion, and the dotted
   line is the fictitious heavy scalar particle.}
\end{figure}

where Roman indices refer to nucleon number, Greek indices refer to
Cartesian coordinates, $\pi(ijk)$ gives all permutations of the indices,
$g_A$ is the axial-vector coupling constant, $F_\pi$ is the pion decay
constant, $\Lambda_\chi$ is taken to be a heavy meson scale, and
$m_\pi$ is the pion mass.
The function $F_{ijk}^{\alpha\beta}$ is defined in Ref.~\cite{tews2016}
and depends on the LECs, $c_1$, $c_3$, and $c_4$. The two LECs $c_D$ and
$c_E$ first appear in the \nnn{} sector at \nxlo{2} and have to be
fitted to $A\ge3$ experimental data.
We discuss our fitting procedure further below.
\subsubsection{Local \texorpdfstring{$\mathit{\nnn{}}$}{3N}
interactions}
The Fourier transformations of Eqs.~(\ref{subeq:qf1}) to (\ref{subeq:qf3}) can
be found in Ref.~\cite{tews2016}. Here, we briefly review some important 
details from that work and point out additional details that arose in the
implementation of the coordinate-space interactions in finite nuclei and 
neutron matter.

In commonly used phenomenological models, any short-range structures
which arise in the Fourier transformation of long-range parts of the
\nnn{} forces are typically absorbed by other short-range structures
(e.g., the scalar short-range structure in the Urbana IX (UIX) \nnn{}
interaction): However, we retain these additional structures explicitly.
Our regularization scheme for the \nnn{} interactions is consistent with
that used in the \nn{} sector, i.e., $\delta$~functions
denoting contact interactions are replaced with
Eq.~(\ref{eq:smeareddelta}),
long-range pion-exchange interactions are regulated by
applying Eq.~(\ref{eq:longrangereg}), and the \nnn{} cutoff parameter is
taken in the same range as the \nn{} cutoff parameter (in the following,
we choose $R_{\nnn{}}=R_0=1.0-1.2$~fm).
The full Fourier transformations of Eq.~(\ref{subeq:qf1}) are available in
Ref.~\cite{tews2016}, but we note that a compact form of
$V_{C,c_3}^{ijk}$ and $V_{C,c_4}^{ijk}$ can be obtained by writing them
in the form of an anticommutator and a commutator of a modified
coordinate-space pion propagator
\begin{equation}
\label{eq:modpionprop}
\mathcal{X}_{ij}(\vb{r})\equiv X_{ij}(\vb{r})-
\frac{4\pi}{m_\pi^2}\drtn{r}\sdots{i}{j}\,.
\end{equation}
See the~\hyperref[app]{Appendix} for details.

\subsubsection{Regulator artifacts}
As was discussed in Refs.~\cite{tews2016, lynn2016, dyhdalo2016}, the
use of local regulators in the \nnn{} sector leads to two kinds of
observable regulator artifacts.
The first kind of regulator artifact affects the short-range parts of the 
interactions in Eqs.~(\ref{subeq:qf2}) and (\ref{subeq:qf3}).
These parts retain additional ambiguities at finite cutoff
$R_{\nnn{}}\ne0$.
The first ambiguity concerns the choice of momentum variables in the
Fourier transformation.
Depending on how this choice is made, Eq.~(\ref{subeq:qf2}) Fourier transforms
to one of the following two equations:
\begin{widetext}
\begin{subequations}
\begin{align}
\label{eq:vdcoord1}
V_{D1}&=
\frac{g_A c_D m_\pi^2}{96 \pi \Lambda_\chi F_\pi^4}
\sum_{i<j<k}\sum_\text{cyc}\tdott{i}{k}\left[
\vphantom{\frac{8\pi}{m_\pi^2}}
X_{ik}(\vb{r}_{kj})\drtn{r_{ij}}
+X_{ik}(\vb{r}_{ij})\drtn{r_{kj}}-\frac{8\pi}{m_\pi^2}\sdots{i}{k}
\drtn{r_{ij}}\drtn{r_{kj}}\right]\,,\\
\label{eq:vdcoord2}
V_{D2}&=
\frac{g_A c_D m_\pi^2}{96 \pi \Lambda_\chi F_\pi^4}
\sum_{i<j<k}\sum_\text{cyc}\tdott{i}{k}\left[
\vphantom{\frac{4\pi}{m_\pi^2}}X_{ik}(\vb{r}_{ik})
-\frac{4\pi}{m_\pi^2}\sdots{i}{k}\drtn{r_{ik}}\right]
\left[\drtn{r_{ij}}+\drtn{r_{kj}}\right]\,,
\end{align}
\end{subequations}
\end{widetext}
where $X_{ik}(\vb{r})=[S_{ik}(\vb{r})T(r)+\sdots{i}{k}]Y_{ik}(r)$ is the
coordinate-space pion propagator, and the tensor and Yukawa
functions are defined as $T(r)=1+3/(m_\pi r)+3/(m_\pi r)^2$ and
$Y(r)=\eu{-m_\pi r}/r$.
The sum with $i<j<k$ runs over all particles 1 to $A$, and the cyclic
sum runs over the cyclic permutations of a given triple.
It is clear that in the limit $R_{\nnn{}}\to0$ the two
possible $V_D$ structures are identical, because then the
$\delta$~functions enforce $i=j$ $(k=j)$ in the first (second) term.
The interaction $V_D$ does not distinguish which of the two nucleons in
the contact interaction participates in the pion exchange.
The term $V_{D2}$ can also be obtained by imagining a heavy fictitious
scalar particle being exchanged between the two nucleons in the contact;
see Fig.~\ref{fig:feyndiagftcd2}.
This ambiguity was already pointed out in Ref.~\cite{navratil2007}.

The second ambiguity in the \nnn{} short-range interactions relates to
the choice of the contact operator in Eq.~(\ref{subeq:qf3}).
The same Fierz-rearrangement freedom that allows for a
selection of local contact operators entering in the \nn{} sector at
\nxlo{1} also allows for the selection of one out of the following six
operators in the \nnn{} sector~\cite{epelbaum2002}:
\begin{equation}
\begin{split}
\{\mathbbm{1},&\sdots{i}{j},\tdott{i}{j},\sdots{i}{j}\tdott{i}{j},\\
&\sdots{i}{j}\tdott{i}{k},
[(\vb*{\sigma}_i\cross\vb*{\sigma}_j)\vdot\vb*{\sigma}_k]
[(\vb*{\tau}_i\cross\vb*{\tau}_j)\vdot\vb*{\tau}_k]\}\,.
\end{split}
\end{equation}
The usual choice is $\tdott{i}{j}$.
This Fierz-rearrangement freedom holds as long as the regulator is
symmetric under individual nucleon permutations.
However, in the presence of local regulators, the Fierz-rearrangement
freedom is violated, and different operator choices can lead to
different results. 
Corrections to the violated Fierz rearrangement freedom are of
higher order in chiral EFT. A systematic study of these effects in 
the \nn{} sector is in preparation~\cite{Huth:2017wzw}.
In the following, we have explored three different choices for the
contact operator:
\begin{subequations}
\begin{align}
\label{eq:vetau}
V_{E\tau}&=\frac{c_E}{\Lambda_\chi F_\pi^4}\sum_{i<j<k}\sum_\text{cyc}
\tdott{i}{k}\drtn{r_{kj}}\drtn{r_{ij}}\,,\\
\label{eq:veone}
V_{E\mathbbm{1}}&=\frac{c_E}{\Lambda_\chi
F_\pi^4}\sum_{i<j<k}\sum_\text{cyc}
\drtn{r_{kj}}
\drtn{r_{ij}}\,,\\
\label{eq:vep}
V_{E\mathcal{P}}&=\frac{c_E}{\Lambda_\chi
F_\pi^4}\sum_{i<j<k}\sum_\text{cyc}\mathcal{P}\,
\drtn{r_{kj}}
\drtn{r_{ij}}\,.
\end{align}
\end{subequations}
The first two operator structures are chosen because $\mathbbm{1}$ 
and $\tdott{i}{j}$ have opposite signs in light nuclei but the same sign
in neutron matter and thus give an estimate of the uncertainty due to
this ambiguity.
The last choice contains the projection operator $\mathcal{P}$ 
that projects on to triples with $S=\tfrac{1}{2}$ and $T=\tfrac{1}{2}$,
\begin{equation}
\label{eq:projector}
\mathcal{P}\equiv\frac{1}{36}\Big(3-\sum_{i<j}\sdots{i}{j}\Big)
\Big(3-\sum_{k<l}\tdott{k}{l}\Big)\,,
\end{equation}
where the sums are over pairs in a given triple. These are the triples
that survive in the limit $\drtn{r}\to\delta^{(3)}(\vb{r})$, that is,
the limit $R_{\nnn{}}\to0$ (or $\Lambda \to\infty$ in momentum space).

The second regulator artifact affects the long-range \nnn{} TPE
interaction.
It has been found that the effective \nnn{} cutoff for
a local regulator is lower (in momentum space) than for a typical
nonlocal regulator~\cite{tews2016, dyhdalo2016}.
As a consequence, one finds less repulsion from a local \nnn{} TPE
interaction than for the standard nonlocal formulation.
This, again, is a regulator artifact that vanishes when $R_{\nnn{}}\to0$.
Lowering the \nnn{} cutoff well below the \nn{} cutoff, however, leads
to collapses because the increasing \nnn{} attraction cannot be
counteracted by additional \nn{} repulsion; see Ref.~\cite{tews2016}.

\subsubsection{Fitting procedure}
We now turn to the fitting procedure for the
LECs $c_D$ and $c_E$.
This procedure was presented and discussed in Ref.~\cite{lynn2016}, but
we review it here for completeness.
In the past, the binding energies of \isotope[3]{H} and \isotope[4]{He}
or the binding energy of \isotope[3]{H} and the $nd$ doublet scattering
length $^2a_{nd}$ have been used to fix $c_D$ and $c_E$.
However, these observables are correlated and thus underconstrain the
two LECs~\cite{gazit2009}.
The \nnn{} couplings have also been fit to the \isotope[3]{H} binding
energy and the \isotope[4]{He} radius~\cite{hebeler2011}.
Arguments can be made that \nnn{} interactions should be fit in $A\le3$
systems only~\cite{gazit2009} or that reproducing observables over a
wider range in the nuclear chart is more
appropriate~\cite{pieper2001,ekstrom2015}.
We take a middle-ground approach and have two goals with 
our fitting strategy: (1) to probe
properties of light nuclei and (2) to probe $T=3/2$ physics.
With these in mind, we take as observables the \isotope[4]{He} binding
energy and \nalpha{} scattering $P$-wave phase shifts.
The \nalpha{} system is the lightest nuclear system for which three
neutrons can be found interacting and thus provides an indirect
constraint on $T=3/2$ physics.

We first find contours for $c_D$ and $c_E$ that reproduce the
\isotope[4]{He} binding energy.
We further constrain $c_D$ and $c_E$ by calculating the $P\ 3/2^-$ and
$P\ 1/2^-$ phase shifts for the \nalpha{} system as described in
Ref.~\cite{nollett2007} and demanding a good reproduction 
of the splitting between these two $P$-wave phase shifts.
See Ref.~\cite{lynn2016} for more details.

In Ref.~\cite{lynn2016}, we explored the various combinations of $V_D$
[Eqs.~(\ref{eq:vdcoord1}) and (\ref{eq:vdcoord2})] and $V_E$
[Eqs.~(\ref{eq:vetau})--(\ref{eq:vep})] and found some
dependence on these choices.
In particular, no fit to both observables (the \isotope[4]{He} binding
energy and the \nalpha{} $P$-wave scattering phase shifts) was obtained
for the case with $V_{D1}$ and the softer cutoff $R_0=1.2$~fm.
For all other combinations, results for light nuclei with $A=3,4$ were
similar.
Below we take a representative choice ($V_{D2},V_{E\tau}$) for the
results we display.

\section{Quantum Monte Carlo methods}
\label{sec:qmc}
In this section, we provide details on QMC methods including the
variational Monte Carlo (VMC) method, which is used as a starting point
for both GFMC and AFDMC calculations.

\subsection{Variational Monte Carlo}
\label{subsec:vmc}
The variational Monte Carlo (VMC) method relies on the Rayleigh-Ritz
variational principle:
\begin{equation}
\label{eq:varprinciple}
\frac{\ev{H}{\Psi_T}}
{\braket{\Psi_T}{\Psi_T}}\geqslant E_0\,,
\end{equation}
where $\ket{\Psi_T}=\ket{\Psi_T(\{c_i\})}$ is a trial wave function with
a set of adjustable parameters $\{c_i\}$, and $E_0$ is the energy of the
ground state of $H$.
The equality above only holds if $\ket{\Psi_T}=\ket{\Psi_0}$,
the ground state of $H$.

\begin{figure*}[ht!]
   \includegraphics[width=2\columnwidth]{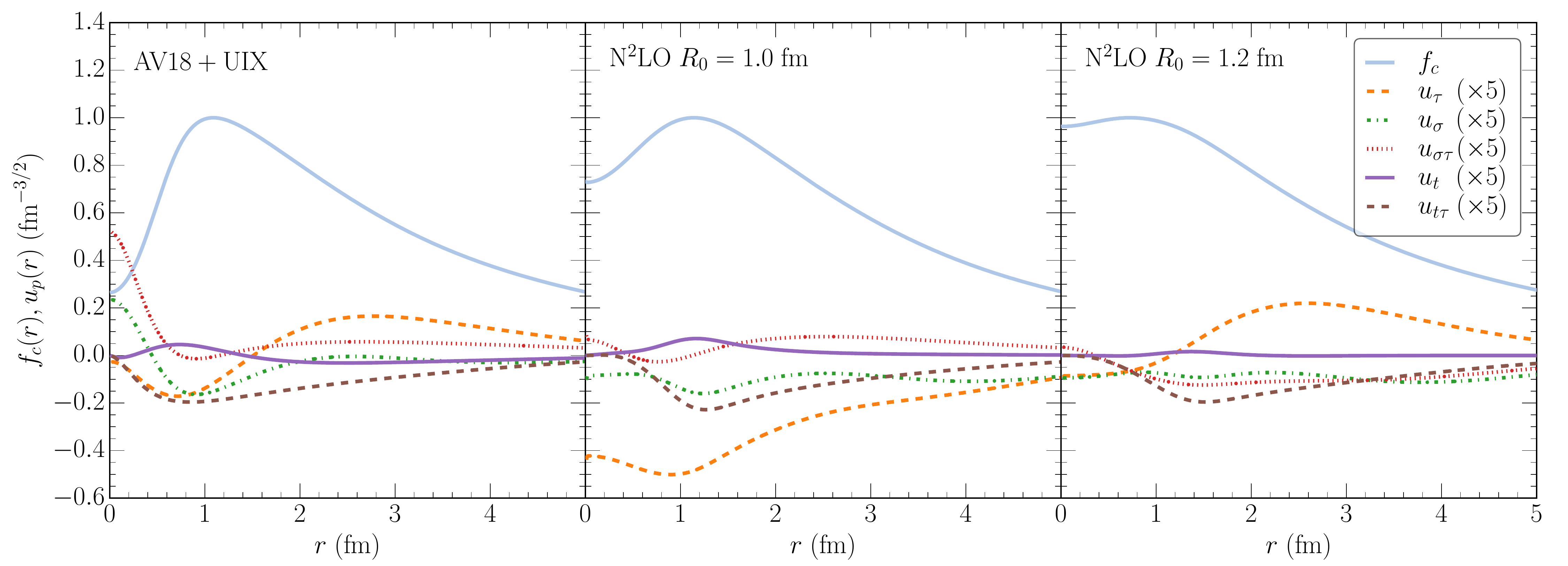}
   \caption{\label{fig:vmccorr}
   Correlations of Eqs.~(\ref{eq:jastrow}) and (\ref{eq:correl})
   entering the trial wave functions used in the calculations of
   $\isotope[4]{He}$ for the $\text{AV18}+\text{UIX}$ (left panel),
   \nxlo{2}~$R_0=1.0$~fm, (middle panel), and \nxlo{2}~$R_0=1.2$~fm
   (right panel) interactions.
   }
\end{figure*}
For few-body nuclei with $A=3,4$ the form of the variational trial
wave function is given as
\begin{equation}
\label{eq:trialwvfun}
\ket{\Psi_T}=\Big[1+\sum_{i<j<k}U_{ijk}\Big]
\Big[\mathcal{S}\prod_{i<j}(1+U_{ij})\Big]\ket{\Psi_J}\,.
\end{equation}
The two-body part of the wave function consists of a symmetrized product
of correlation operators acting on a Jastrow wave function,
\begin{equation}
\label{eq:2ntrialwvfun}
\Big[\mathcal{S}\prod_{i<j}(1+U_{ij})\Big]\ket{\Psi_J}\,,
\end{equation}
where the Jastrow wave function is
\begin{equation}
\label{eq:jastrow}
\ket{\Psi_J}=\prod_{i<j}f_c(r_{ij})\ket{\Phi}\,.
\end{equation}
The Jastrow factor is a product of central two-body correlations $f_c$ acting
on an appropriate antisymmetric single-particle state.
For few-body nuclei with $A=3,4$, $\ket{\Phi}$ can be taken as an
appropriate antisymmetric linear combination of spin-isospin states.
For example, for \isotope[4]{He}, one can take
\begin{equation}
\ket{\Phi_4}=\mathcal{A}\ket{p\!\up p\!\down n\!\up n\!\down}\,.
\end{equation}
The correlation operators are defined as
\begin{equation}
\label{eq:correl}
U_{ij}=\sum_{p}u_p(r_{ij})O^{(p)}_{ij}\,,
\end{equation}
where the $\{O_{ij}^{(p)}\}$ are the operators
\begin{equation}
\left\{\sdots{i}{j},\tdott{i}{j},\sdots{i}{j}\tdott{i}{j},
S_{ij},S_{ij}\tdott{i}{j},\vb{L}\vdot\vb{S}\right\}\,,
\end{equation}
taken from the two-body interaction.
We use the short-hand notation $p=\{\sigma,\tau,\sigma\tau,t,t\tau,b\}$
for the operators as in Ref.~\cite{wiringa1991}.
The symmetrizer in Eqs.~(\ref{eq:trialwvfun}) and (\ref{eq:2ntrialwvfun})
is necessary to maintain the overall antisymmetry of the wave function,
since in general the $U_{ij}$ do not commute with each other.
In Fig.~\ref{fig:vmccorr}, we display the two-body correlations $f_c$,
and $\{u_p\}$ obtained in the simulation of \isotope[4]{He} with the
\nxlo{2} interactions with both cutoffs as well as those obtained for
the Argonne $v_{18}$ \nn{} interaction supplemented by the UIX \nnn{}
interaction.
What can be seen from these correlations, most particularly in the case
of the central correlation $f_c$, is the softening of the interaction as
we take the cutoff from $R_0=1.0$~fm to $R_0=1.2$~fm.
We find that the spin-orbit correlation has only a minimal effect on the
variational energies we obtain and a relatively high computational cost,
and therefore we set $u_b(r_{ij})=0$ in our calculations.

The three-body correlation operator takes the following form:
\begin{equation}
\label{eq:3ncorr}
U_{ijk}=\epsilon V_{ijk}(\bar{r}_{ij},\bar{r}_{jk},\bar{r}_{ik})\,,
\end{equation}
where $\bar{r}$ is a scaled relative separation and $\epsilon$ is a
small negative constant.
This form is suggested by perturbation theory~\cite{pudliner1997}.
In addition to the explicit three-body correlations of
Eq.~(\ref{eq:3ncorr}), a central, geometric three-body correlation is
wrapped into the two-body correlations,
\begin{equation}
\tilde{u}_p(r_{ij})=\prod_{k\ne i,j}f_{ijk}u_p(r_{ij})\,,
\end{equation}
with
\begin{equation}
f_{ijk}=1-t_1\!\left(\frac{r_{ij}}{R_{ijk}}\right)^{t_2}
\!\!\exp(-t_3R_{ijk})\,,
\end{equation}
where $R_{ijk}=r_{ij}+r_{jk}+r_{ik}$ and the $\{t_i\}$ are variational
parameters.
These correlations serve to reduce the repulsion which arises from
the product of certain spin-isospin correlation operators when any two
nucleons come close together.
Reducing this repulsion was found to improve variational energies with
wave functions of the form
of Eq.~(\ref{eq:2ntrialwvfun})~\cite{lomnitz-adler1981}.

Equation~(\ref{eq:varprinciple}) is evaluated by means of Monte Carlo
integration,
\begin{equation}
\label{eq:vmc}
\ev{H}=\frac{\sum_{a,b}\int\dd\vb{R}[\Psi_a^\dagger(\vb{R})H
\Psi_b(\vb{R})/W_{ab}(\vb{R})]W_{ab}(\vb{R})}
{\sum_{a,b}\int\dd\vb{R}[\Psi_a^\dagger(\vb{R})
\Psi_b(\vb{R})/W_{ab}(\vb{R})]W_{ab}(\vb{R})}\,,
\end{equation}
where $a$ and $b$ stand for a given order of operators in the
product Eq.~(\ref{eq:2ntrialwvfun}),
a complete sum over all spin and isospin states is assumed, and the
integrals are performed as a Monte Carlo integration over the
coordinate-space configurations
$\vb{R}=\{\vb{r}_1,\vb{r}_2,\ldots,\vb{r}_A\}$.
The sums over the orders $a$ and $b$ are also performed via a Monte
Carlo sampling as discussed below.
The weight function can be taken as
\begin{equation}
\label{eq:simpweight}
W_{ab}(\vb{R})=|\Re{\langle\Psi_a^\dagger(\vb{R})\Psi_b(\vb{R})
\rangle}|\,,
\end{equation}
for example.
In practice, because of the different orders $a$ and $b$ in the left and
right wave functions, Eq.~(\ref{eq:simpweight}) is not guaranteed to be
nonzero, and so we add to it an additional term proportional to
$\sum_{s,t}|\Psi_{a}(\vb{R};s,t)^\dagger\Psi_b(\vb{R};s,t)|$. 
That is, we add a term proportional to the sum of the absolute value of
the overlaps of the individual spin-isospin components of the wave
functions.

The symmetrizer in Eqs.~(\ref{eq:trialwvfun}) and
(\ref{eq:2ntrialwvfun}) requires, in principle, the evaluation of all
$[A(A-1)/2]!$ possible orderings of the operators.
To save computational cost, the order of operators is instead sampled.
This approximation does not contribute much to the statistical variance
since all orderings share the same linear (dominant) contributions and
the differences between different orderings are proportional to
$\{u_p^2\}$.

The Metropolis algorithm is employed and the result is, after sufficient
equilibration, a set of configurations labeled by the $3A$ coordinates
and the orderings of the operators, $\{\vb{R},a,b\}$, which are
distributed according to the square of the trial wave function.
As the integration and sum over all orderings is done stochastically,
there is an error associated with the expectation value of any operator
$\ev{O}$, given by
\begin{equation}
\sigma_O=\sqrt{\frac{\ev{O^2}-\ev{O}^2}{N-1}}\,,
\end{equation}
where $N$ is the number of statistically independent evaluations.
For more details, see Refs.~\cite{wiringa1991,carlson2015}.

With the algorithm described above, the variational parameters $\{c_i\}$
are adjusted to minimize the expectation value of the
Hamiltonian in Eq.~(\ref{eq:varprinciple}).
Wave functions so obtained can be used as reasonable approximations to
the exact ground state (especially in few-body nuclei) and are a
necessary starting point for the GFMC method.

\subsection{Diffusion Monte Carlo}
Even with the sophisticated wave functions described
in Sec.~\ref{subsec:vmc}, it is not possible to construct by hand exact
solutions to the many-body Schr\"odinger equation.
Diffusion Monte Carlo methods including the AFDMC and GFMC methods rely
on the fact that, given a nuclear system specified by the Hamiltonian
$H$ with ground state $\ket{\Psi_0}$ and a trial wave function for that
system $\ket{\Psi_T}$ with nonvanishing overlap with the ground state, 
\begin{equation}
\lim_{\tau\to\infty}\eu{-H\tau}\ket{\Psi_T}\to\ket{\Psi_0}\,.
\end{equation}
The object $\eu{-H\tau}$ is the many-body imaginary-time Green's
function (or imaginary-time propagator) for the system with the
imaginary time $\tau$.
This ``sifting'' property of the imaginary-time propagator is easy to
understand if the trial wave function is expanded in a complete set of
eigenstates of $H$, $\{\ket{\phi_n}\}$, with energies $\{E_n\}$,
\begin{equation}
\label{eq:dmcexpand}
\eu{-(H-E_T)\tau}\ket{\Psi_T}=
\sum_{n=0}^\infty\eu{-(E_i-E_T)\tau}a_n\ket{\phi_n}\,,
\end{equation}
where we have introduced the trial energy $E_T$ and
$a_n=\langle\phi_n|\Psi_T\rangle$.
In principle, $E_T$ can take any value, but it is often adjusted to be
the ground-state energy (or the energy of the low-lying excited state
sought).
Then, since $E_i>E_T$ for all $i>0$, in the large-imaginary-time limit
all of the excited-state components of the trial state are exponentially
damped and one is left with the exact many-body ground state.
In this language, we can say that with the VMC method alone it is not
possible to avoid some contamination in nuclear wave functions from
excited states.
That is, while we can make $a_0$ of Eq.~(\ref{eq:dmcexpand}) the dominant
contribution through the adjustment of the variational parameters
$\{c_i\}$, it is not possible with the VMC method alone to guarantee
that $a_{n>0}=0$.

In the remainder of this section, we discuss diffusion Monte Carlo
methods, paying particular attention to the GFMC method, which we 
use to calculate properties of light nuclei.
For more details, we refer to Ref.~\cite{carlson2015} and references
therein.
We begin with a discussion of the calculation of the imaginary-time
propagator, which plays a central role in diffusion Monte Carlo methods.

In general, it is difficult to compute the exact many-body
imaginary-time propagator for arbitrary imaginary times.
Instead, the properties of the exponential are exploited to
rewrite the propagation to large imaginary time as a product of small
propagations,
\begin{equation}
\eu{-H\tau}=\prod_{i=1}^N\eu{-H\Delta\tau}\,,
\end{equation}
with $\Delta\tau=\tau/N$, and $N$ large enough ($\Delta\tau$ small
enough) such that one of several approximations can be used to calculate
the short-imaginary-time propagator.
In the case of the AFDMC method, for example, a Trotter
breakup is used~\cite{carlson2015},
\begin{equation}
\label{eq:trot}
\eu{-H\Delta\tau}=
\bigg[\prod_{i<j}\eu{-V_{ij}\frac{\Delta\tau}{2}}\bigg]
\eu{-T\Delta\tau}
\bigg[\prod_{i<j}\eu{-V_{ij}\frac{\Delta\tau}{2}}\bigg]
+\mathcal{O}(\Delta\tau^3)\,,
\end{equation}
where $T$ is the kinetic energy operator and $V_{ij}$ is a local
two-body interaction.
In Eq.~(\ref{eq:trot}), the order in the product on the left is taken in 
the opposite order of the product on the right.
This keeps the propagator unitary (in real time) and eliminates terms of
$\mathcal{O}(\Delta\tau^2)$.

In the GFMC method, the exact two-body propagator is used to construct
the many-body propagator, as suggested by studies of condensed helium
systems~\cite{ceperley1995}:
\begin{equation}
\begin{split}
&\matrixel{\alpha\vb{R}}{\eu{-H\Delta\tau}}{\beta\vb{R}^\prime}\\
&\equiv G_{\alpha\beta}(\vb{R},\vb{R}^\prime;\Delta\tau)\\
&=G_0(\vb{R},\vb{R}^\prime;\Delta\tau)
\Big\langle\alpha
\Big|\mathcal{S}\prod_{i<j}
\frac{g_{ij}(\vb{r}_{ij},\vb{r}_{ij}^\prime;\Delta\tau)}
{g_{0,ij}(\vb{r}_{ij},\vb{r}_{ij}^\prime;\Delta\tau)}\Big|
\beta\Big\rangle+\mathcal{O}(\Delta\tau^3)\,.
\end{split}
\end{equation}
Here, $\alpha$ and $\beta$ stand for the appropriate spin-isospin states
for a given nucleus, $\vb{R}=\{\vb{r}_1,\vb{r}_2,\ldots,\vb{r}_A\}$
and
$\vb{R}^\prime=\{\vb{r}^\prime_1,\vb{r}_2^\prime,\ldots,
\vb{r}^\prime_A\}$ are the collections of $3A$ coordinates before and
after the propagation step, $G_0(\vb{R},\vb{R}^\prime;\Delta\tau)$ is
the many-body free-particle imaginary-time propagator
\begin{equation}
G_0(\vb{R},\vb{R}^\prime;\Delta\tau)=
\left(\frac{m}{2\pi\hbar^2\Delta\tau}\right)^{\frac{3A}{2}}
\exp
\left[-\frac{(\vb{R}-\vb{R}^\prime)^2}{2\hbar^2\Delta\tau/m}\right]\,,
\end{equation}
$g_{ij}$ is the exact two-body interacting imaginary-time propagator,
\begin{equation}
\label{eq:xct2bdyprop}
g_{ij}(\vb{r}_{ij},\vb{r}_{ij}^\prime;\Delta\tau)=
\matrixel{\vb{r}_{ij}}{\eu{-H_{ij}\Delta\tau}}{\vb{r}_{ij}^\prime}\,,
\end{equation}
which can be computed to high accuracy ($\sim$8- to 10-digit accuracy or
better than half machine precision),
and $g_{0,ij}$ is the two-body free-particle analog of $g_{ij}$.
This construction allows for taking much larger time steps than in the
Trotter breakup in Eq.~(\ref{eq:trot}).
The trade-off is that the calculation of the exact two-body
propagator of Eq.~(\ref{eq:xct2bdyprop}) is too costly to compute ``on the fly''
and must be carried out in advance and stored on a grid of points to be
interpolated on during the GFMC propagation.

The complete two-body propagator depends on initial and final relative
coordinates, the initial and final spin states of the pair, and the
isospin of the pair,
\begin{equation}
\begin{split}
&\matrixel{\alpha}{g(\vb{r},\vb{r}^\prime;\Delta\tau)}
{\beta}\\
&\ \ \ \to\matrixel{\vb{r}^\prime SM_S^\prime T M_T}
{\eu{-H\Delta\tau}}{\vb{r}SM_STM_T}\,,
\end{split}
\end{equation}
where the indices $ij$ as in Eq.~(\ref{eq:xct2bdyprop}) are suppressed
here and in what follows for simplicity unless they are explicitly
needed for clarity.
Reference~\cite{schmidt1995} originally proposed using fast Fourier 
transforms (FFT) and the Trotter break up for scalar interactions, and
this idea was adapted to realistic nuclear interactions in 
Ref.~\cite{pudliner1997}.
In this method, interactions are first decomposed into partial
waves. The nuclear Hamiltonian commutes with the operators $J^2$,
$J_z$, $S^2$, $T^2$, and $T_z$, and, thus, sets them as good 
channel quantum numbers:
$\vb{S}=\vb{S}_1+\vb{S}_2$ is the total spin,
$\vb{J}=\vb{L}+\vb{S}$ is the total angular momentum, and
$\vb{T}=\vb{T}_1+\vb{T}_2$ is the total isospin.
Then, the channel propagators $\matrixel{r^\prime JM_JL^\prime S TM_T}
{\eu{-H\Delta\tau}}{rJM_JLSTM_T}$ are computed and resummed to
obtain the two-body propagator:
\begin{equation}
\begin{split}
&\matrixel{\vb{r}^\prime SM_S^\prime TM_T}{\eu{-H\Delta\tau}}
{\vb{r}SM_STM_T}\\
&=\sum_\gamma C_{SM_S^\prime L^\prime M_L^\prime}^{JM}
Y_{L^\prime M_L^\prime}(\Omega^\prime)C_{SM_SLM_L}^{JM}
Y_{LM_L}^\ast(\Omega)\\
&\times\matrixel{r^\prime JM_JL^\prime S TM_T}
{\eu{-H\Delta\tau}}{rJM_JLSTM_T}\,.
\end{split}
\end{equation}
Here, $\gamma$ stands for the set of quantum numbers $\{JMLL^\prime
M_LM_L^\prime\}$, $C$ is a Clebsch-Gordan coefficient, $Y$ is a
spherical harmonic, and $\Omega$ ($\Omega^\prime$) are the angular
coordinates of $\vb{r}$ ($\vb{r}^\prime$).

Each of the channel propagators is calculated by breaking up the
(already-small) time step $\Delta\tau$ into smaller steps
$\delta\tau=\Delta\tau/N_\tau$, with $N_\tau$ large, using the
symmetrized Trotter breakup, and FFT: 
\begin{subequations}
\begin{align}
\eu{-H\Delta\tau}&=
(\eu{-H\delta\tau})^{N_\tau}\,,\\
\label{eq:chanprop2}
\eu{-H\delta\tau}&=
\eu{-V\delta\tau/2}\eu{-T\delta\tau}\eu{-V\delta\tau/2}+
\mathcal{O}(\delta\tau^3)\,.
\end{align}
\end{subequations}
In Eq.~(\ref{eq:chanprop2}), the right-most exponential acts upon an array of
initial relative separations, the result is transformed to momentum
space using FFT, the exponential of the kinetic energy acts upon that
result, which is then transformed back to coordinate space using FFT,
whereupon the left-most exponential acts upon the array.
This method introduces errors of $\mathcal{O}(\delta\tau^3)$, is fast,
and is easy to implement.

An alternative method is to diagonalize the channel Hamiltonians in
momentum space~\cite{lynn2012}.
When the interaction is nonlocal (no longer diagonal in coordinate
space), then the advantages of the Trotter breakup vanish.
That is, it is just as difficult to calculate the matrix elements
$\matrixel{\vb{r}^\prime}{\eu{-V\Delta\tau}}{\vb{r}}$ as it
is to calculate the original matrix elements
$\matrixel{\vb{r}^\prime}{\eu{-H\Delta\tau}}{\vb{r}}$.
In order to diagonalize the channel Hamiltonians, we take as an
orthonormal basis the set of spherical Bessel functions which solve the
free radial Schr\"odinger equation with a Dirichlet boundary condition
at some radius $R$ much beyond the range of the interaction, 
\begin{equation}
\phi_{nL}(r)=
\sqrt{\frac{2}{R^3j_L^\prime(k_nR)^2}}j_L(k_nr)\,,
\end{equation}
where $\{k_n\}$ is the set of discrete momenta for a given $L$ and
$R$.
In this basis, the kinetic energy is diagonal, and the potential-energy
matrix elements can be obtained with simple matrix multiplications, which
perform the necessary numerical integrals.
While this method was originally developed to calculate two-body
propagators for nonlocal interactions, it works equally well for local
interactions, providing equal accuracy and speed when compared with the
symmetrized Trotter break up with FFT.

So far, we have discussed only the contribution to the many-body
propagator coming from \nn{} interactions.
We include \nnn{} interactions in the propagator as a symmetric
linear approximation to $\eu{-V_{\nnn{}}\Delta\tau}$:
\begin{widetext}
\begin{equation}
\begin{split}
G_{\alpha\beta}(\vb{R},\vb{R}^\prime;\Delta\tau)=
G_0(\vb{R},\vb{R}^\prime;\Delta\tau)&
\Big\langle\alpha
\Big|\mathbbm{1}-\frac{\Delta\tau}{2}\sum_pV^{(p)}_{\nnn{}}(\vb{R})
\Big|\gamma\Big\rangle\\
&\times\Big\langle\gamma\Big|\mathcal{S}\prod_{i<j}
\frac{g_{ij}(\vb{r}_{ij},\vb{r}_{ij}^\prime;\Delta\tau)}
{g_{0,ij}(\vb{r}_{ij},\vb{r}_{ij}^\prime;\Delta\tau)}
\Big|\delta\Big\rangle
\Big\langle\delta
\Big|\mathbbm{1}-
\frac{\Delta\tau}{2}\sum_pV^{(p)}_{\nnn{}}(\vb{R}^\prime)
\Big|\beta\Big\rangle\,,
\end{split}
\end{equation}
\end{widetext}
where the sums $\sum_pV^{(p)}_{\nnn{}}$ are over all \nnn{}
operators of Eqs.~(\ref{eq:compvcc1})--(\ref{eq:compvcc4}), one of
Eqs.~(\ref{eq:appvdcoord1}) and (\ref{eq:appvdcoord2}), and one of
Eqs.~(\ref{eq:vetdott})--(\ref{eq:veproj}).
As before, $\alpha$, $\beta$, $\gamma$, and $\delta$ are appropriate
spin-isospin states and $\gamma$ and $\delta$ are summed over.
This linear approximation is a controlled approximation that becomes
more exact with smaller $\Delta \tau$.
There are improvements to this linear approximation possible.
For example, replacing
$[\mathbbm{1}-\tfrac{\Delta\tau}{2}\sum_pV^{(p)}_{\nnn}(\vb{R})]$ 
with
$\prod_p[\mathbbm{1}-\tfrac{\Delta\tau}{2}V^{(p)}_{\nnn}(\vb{R})]$
would capture at least some $\mathcal{O}(\Delta\tau^2)$ effects.
Another possibility is to include all parts of the \nnn{} interaction
that can be rewritten effectively as two-body operators into the
two-body propagator as suggested in Ref.~\cite{wiringa2000} [these
include the TPE $P$-wave anticommutator contribution
Eq.~\eqref{eq:compvcc3}, the TPE $S$-wave contribution
Eq.~\eqref{eq:compvcc1}, the $V_D$ contributions Eqs.~\eqref{eq:appvdcoord1}
and \eqref{eq:appvdcoord2}, and two of the three $V_E$ contributions
Eqs.~\eqref{eq:vetdott} and \eqref{eq:veoneone}].
However, we have found that with the time step we typically use,
$\Delta\tau=0.0005\text{ MeV}^{-1}$, the time-step error introduced by
this linear approximation is negligible.

With the imaginary-time propagator so obtained, one would ideally like
to calculate expectation values such as
$\ev{O(\tau)}=\tfrac{\matrixel{\Psi(\tau)}{O}{\Psi(\tau)}}
{\braket{\Psi(\tau)}{\Psi(\tau)}}$,
with $\Psi(\tau)$ defined as
\begin{equation}
\Psi(\vb{R}_N;\tau)\equiv\int\prod_{i=0}^{N-1}\dd\vb{R}_i
G(\vb{R}_{i+1},\vb{R}_i;\Delta\tau)\Psi_T(\vb{R}_0)\,.
\end{equation}
However, in practice, one does not have direct access to the 
propagated wave function, and an evaluation of that expectation 
value is cumbersome for spin- and isospin-dependent operators, and
especially for momentum-dependent operators.
Thus, what is more commonly used is the mixed expectation value of a
given operator (suppressing the spin-isospin indices), defined as
\begin{widetext}
\begin{equation}
\label{eq:mixedest}
\ev{O}_\text{mixed}\equiv\frac{\matrixel{\Psi_T}{O}{\Psi(\tau)}}
{\braket{\Psi_T}{\Psi(\tau)}}=\frac{\int\dd\vb*{\mathcal{R}}
\Psi_T^\dagger(\vb{R}_N)OG(\vb{R}_N,\vb{R}_{N-1};\Delta\tau)
\cdots G(\vb{R}_1,\vb{R}_0;\Delta\tau)\Psi_T(\vb{R}_0)}
{\int\dd\vb*{\mathcal{R}}\Psi_T^\dagger(\vb{R}_N)
G(\vb{R}_N,\vb{R}_{N-1};\Delta\tau)\cdots
G(\vb{R}_1,\vb{R}_0;\Delta\tau)\Psi_T(\vb{R}_0)}\,,
\end{equation}
\end{widetext}
with the paths $\dd\vb*{\mathcal{R}}\equiv\prod_{i=0}^{N-1}\dd\vb{R}_i$,
and the total imaginary time $\tau=N\Delta\tau$.
The paths are Monte Carlo sampled to perform the integrals.
Note that the operator $O$ must act on the trial wave function (to the
left).

The mixed estimate introduces an explicit dependence on the trial wave
function. However, if the trial wave function is a good approximation, 
we can write
\begin{equation}
\Psi(\tau)=\Psi_T+\delta\Psi(\tau)\,,
\end{equation}
where $\delta\Psi(\tau)$ is the (small) correction to the trial wave
function introduced by the imaginary-time propagation, and keep terms
only of $\mathcal{O}\left(\delta\Psi(\tau)\right)$.
Then we have
\begin{equation}
\begin{split}
\ev{O(\tau)}&=\frac{\matrixel{\Psi(\tau)}{O}{\Psi(\tau)}}
{\braket{\Psi(\tau)}{\Psi(\tau)}}\\
&\approx\ev{O(\tau)}_\text{mixed}+
\big(\ev{O(\tau)}_\text{mixed}-\ev{O}_T\big)\,,
\end{split}
\end{equation}
where $\ev{O}_T$ is the variational estimate.
Thus, if $\Psi_T$ is a good approximation to the exact wave function
obtained through imaginary-time propagation (as measured by the relative
smallness of the difference
$\ev{O(\tau)}_\text{mixed}-\ev{O}_T$) when compared with
$\ev{O(\tau)}_\text{mixed}$, then the mixed estimate
introduces only a small systematic uncertainty.
Typically we aim for the difference between the mixed and variational
estimates to be no larger than $\sim5\%$ of the mixed estimate.
There are other ways to avoid the use of mixed estimates, such as
computing the observable in the midpoint of the path~\cite{kalos1967},
but this requires a propagation time twice as long as in the
mixed-estimate case.
Note that in the case of the energy expectation value, $\ev{H}$, the
Hamiltonian and the imaginary-time propagator commute.
In this case,
\begin{equation}
\begin{split}
\ev{H(\tau)}_\text{mixed}&=
\frac{\matrixel{\Psi_T}{\eu{-H\tau}H}{\Psi_T}}
{\matrixel{\Psi_T}{\eu{-H\tau}}{\Psi_T}}\\
&=\frac{\matrixel{\Psi_T}{\eu{-H\tau/2}H\eu{-H\tau/2}}{\Psi_T}}
{\matrixel{\Psi_T}{\eu{-H\tau/2}\eu{-H\tau/2}}{\Psi_T}}\\
&=\frac{\matrixel{\Psi(\tau)}{H}{\Psi(\tau)}}
{\braket{\Psi(\tau)}{\Psi(\tau)}}\,,
\end{split}
\end{equation}
such that $\lim_{\tau\to\infty}\ev{H(\tau)}_\text{mixed}=E_0$.
In short, for the Hamiltonian, the mixed estimate is identical to the
normal estimate.

When performing the propagation, one has to employ another 
approximation. Nucleons are fermions and their many-body wave
functions contain nodal surfaces. As a consequence, a configuration 
that crosses a nodal surface introduces a sign change in the matrix
elements in Eq.~\eqref{eq:mixedest}. At large $\tau$, these sign 
changes contribute to a decreasing denominator, causing large 
statistical fluctuations (large variance). This is the famous fermion sign
problem. One way to circumvent this problem is the so-called 
constrained path algorithm; for a detailed description, see
Ref.~\cite{wiringa2000}. In short, the idea is to discard configurations 
that in future propagations would only contribute to the variance.
If one knew the exact wave function, then the overlap of these 
discarded configurations with the ground-state wave function would 
be zero $\braket{\Psi_\text{discarded}}{\Psi_0}=0$. However, since 
we do not in general know the exact ground-state wave function, the
constraint is imposed on the overlap with the trial wave function so 
that the average of the overlaps $\braket{\Psi_\text{discarded}}{\Psi_T}$ 
over the random walk is approximately zero.This approximation was 
inspired by the fixed-node approximation used 
in condensed matter systems.

\begin{figure}[t!]
   \includegraphics[width=\columnwidth]{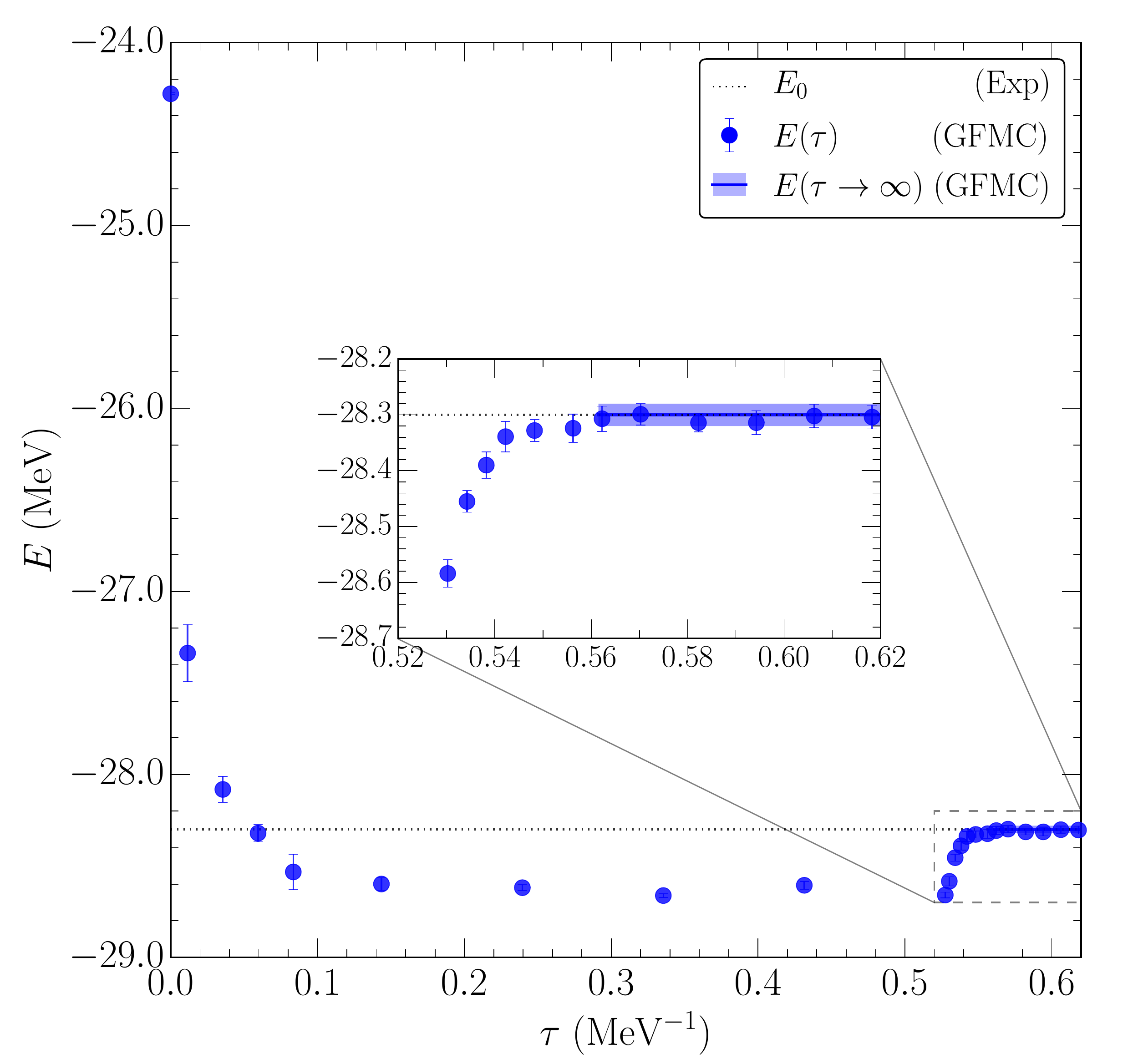}
   \caption{\label{fig:he4cpeandtevstau}
   Energy of \isotope[4]{He} as a function of imaginary time in 
   constrained-path GFMC calculations. Past 
   $\tau \sim$0.5 $\text{MeV}^{-1}$ we show the transient estimation.
   The inset shows the details of the transient estimation and the
   region used to extract the ground-state energy and uncertainty (light
   blue band).
   Note that each point represents an average over a given (varying)
   imaginary-time interval.
   The imaginary-time intervals averaged over are shorter at the
   beginning and end of the propagation in order to show more detail in
   these intervals.
   }
\end{figure}

For scalar wave functions (no spin or
isospin dependence) the fixed-node approximation provides both a way 
to tame the sign problem, and results in an upper bound to the 
ground-state energy. However, because of the spin and isospin
dependence of the nuclear case, the constrained-path algorithm no 
longer supplies a strict upper bound, as has been discussed and
demonstrated in Ref.~\cite{wiringa2000}. To overcome this additional
difficulty, in cases where the constrained-path algorithm is used, we 
take a number $n_u$ of unconstrained steps after convergence of the
constrained-path calculation. We take $n_u$ as large as possible. 
Typically, $n_u \sim20$ before the fermion sign problem overwhelms 
the signal. This ``transient estimation'' results in significantly
improved estimates, introducing an error, for example, in
\isotope[6]{Li} of just $\sim0.5\%$; see Ref.~\cite{wiringa2000}.
Figure~\ref{fig:he4cpeandtevstau} gives an example of a constrained-path
calculation of the ground-state energy of \isotope[4]{He} and the
subsequent transient estimation.
Note that the constrained-path propagation overbinds the system,
demonstrating that for some trial wave functions the constrained-path
estimate is not an upper bound.

\section{Energies and other results for
\texorpdfstring{$\boldsymbol{A=3,4}$}{A=3,4}}
\label{sec:enrgrslts}
\begin{figure*}[htb!!]
   \includegraphics[width=2\columnwidth]{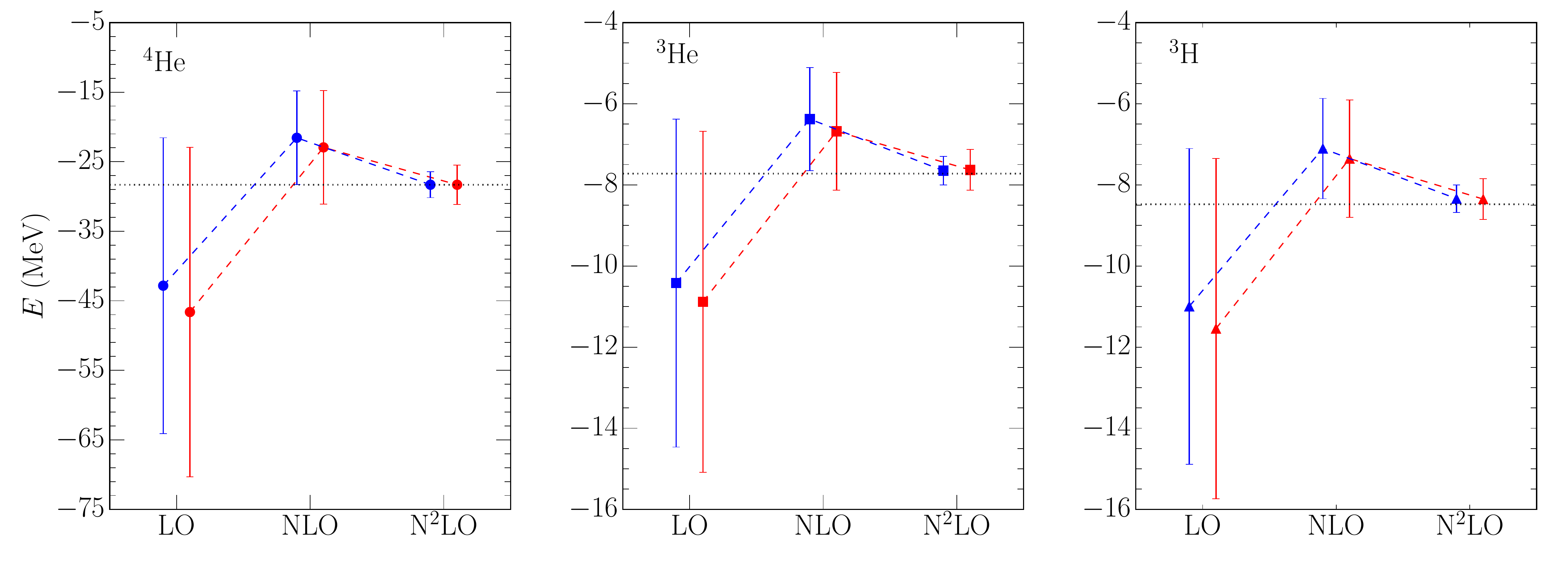}
   \caption{\label{fig:ekme}
   Energies as calculated using the GFMC method at \nxlo{0}, \nxlo{1},
   and \nxlo{2} for $A=3,4$ nuclei.
   The uncertainties include an estimate for the uncertainty coming from
   the truncation of the chiral expansion.
   In blue (red) are the energies with the cutoff $R_0=1.0$~fm
   ($R_0=1.2$~fm).
   The horizontal lines are the experimental values.}
\end{figure*}
\begin{figure*}[htb!]
   \includegraphics[width=2\columnwidth]{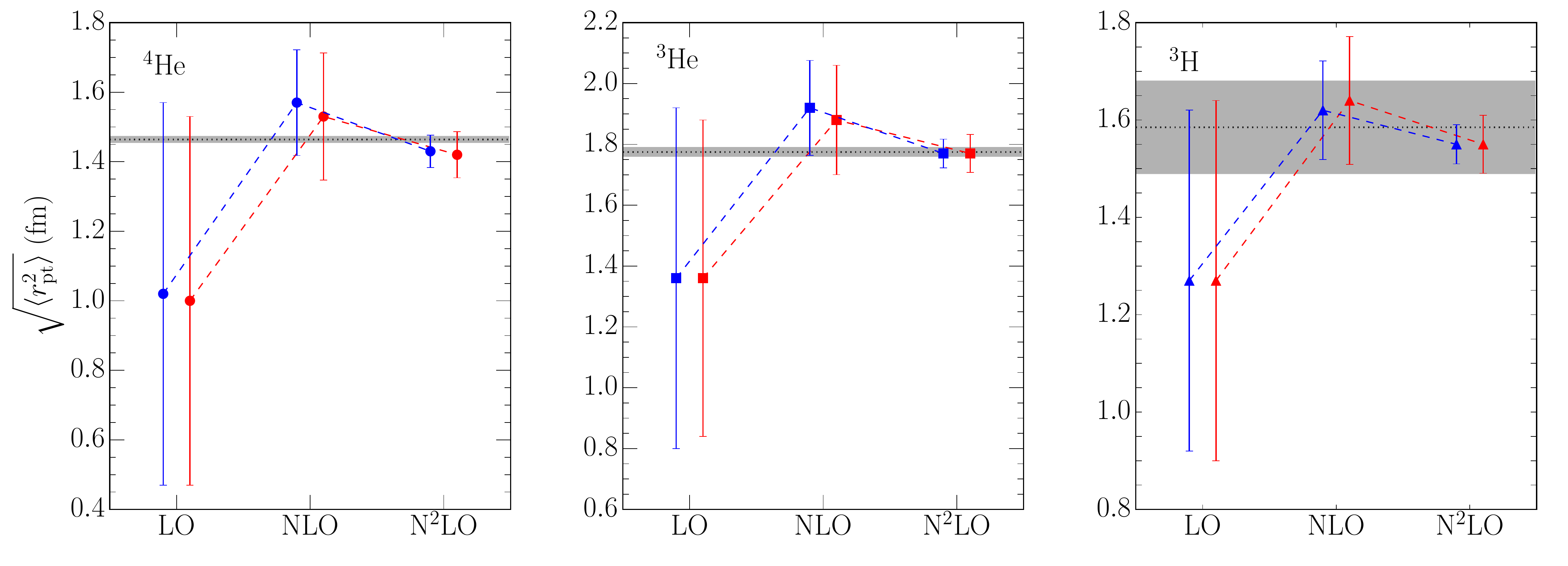}
   \caption{\label{fig:ekmr}
   Point-proton radii as calculated using the GFMC method at \nxlo{0},
   \nxlo{1}, and \nxlo{2} for $A=3,4$ nuclei.
   The uncertainties include an estimate for the uncertainty coming from
   the truncation of the chiral expansion.
   In blue (red) are the energies with the cutoff $R_0=1.0$~fm
   ($R_0=1.2$~fm).
   The horizontal bands are the experimental values with uncertainties.}
\end{figure*}
The light nuclei with $A=3,4$ are a minimal testing ground for any
nuclear Hamiltonian: The reasonable reproduction of binding energies 
and radii in these nuclei is a basic yardstick against which our local 
chiral interactions can be measured.

In this section, we present the main results for light nuclei that we
have obtained with our local chiral EFT \nn{} and \nnn{} interactions at
\nxlo{2}~\cite{lynn2016}.
We emphasize the order-by-order convergence of observables in the
$A=3,4$ nuclei and present a detailed breakdown of the contributions to
the energies from various components of the \nn{} and \nnn{}
interactions.
We also show several one-body distributions and the related longitudinal
charge form factor.

\subsection{Energies of light nuclei at
\texorpdfstring{\nxlo{0}}{LO}, \texorpdfstring{\nxlo{1}}{NLO}, and
\texorpdfstring{\nxlo{2}}{N2LO}}
At \nxlo{0}, the \nn{} interaction consists simply of the one-pion
exchange potential and two contact interactions with LECs fit to \nn{}
scattering phase shifts.
With only the basic pion physics present and little freedom to fit the
phase shifts, essentially only the large scattering length plus OPE 
physics can be
reproduced, and the resulting potential is excessively attractive in
low partial waves.
This can be seen in Fig.~\ref{fig:ekme}, where at \nxlo{0}, the
ground-state energies of $A=3,4$ nuclei are significantly lower than
experiment.
In fact, the \nxlo{0} \nn{} interaction overbinds by as much as
$\sim50$\% ($\sim30$\%) for $A=4$ ($A=3$).
At \nxlo{1}, the \nn{} interaction is too repulsive and leads to 
underbinding. However, the deviation from experiment decreases to 
$\sim25$\% ($\sim15$\%) for $A=4$ ($A=3$).
Finally, at \nxlo{2}, the \nnn{} interaction with two free LECs enters.
We fit $c_D$ and $c_E$ directly to the binding energy of
\isotope[4]{He}, (see Sec.~\ref{subsec:nnn}) and, since the binding
energies of the $A=3$ systems are highly correlated with the binding
energy of \isotope[4]{He} (i.e., the Tjon
line~\cite{tjon1976,platter2005}), the $A=3$ binding energies are also
well reproduced.

The uncertainties in Fig.~\ref{fig:ekme} contain contributions from the
GFMC statistical uncertainties as well as from an estimate for the
theoretical uncertainty coming from the truncation of the chiral
expansion (as discussed in Sec.~\ref{sec:localcheft}).
The theoretical uncertainties display at least three desirable features:
(1) They encompass, order by order, the cutoff variation in the energy,
(2) order by order, the experimental energy is within the uncertainty
bands, and
(3) as the chiral order increases, the uncertainty coming from the
truncation of the chiral expansion decreases rapidly.
Thus, at \nxlo{3}, we can expect that while the energies of these
systems will not change dramatically, the uncertainties will continue to
reduce.

In addition to Fig.~\ref{fig:ekme}, we show more details of various
contributions to the $A=3,4$ energies in GFMC calculations in
Table~\ref{tab:energies}, where the softer nature of the interaction
with $R_0=1.2$~fm is evident from the lower kinetic energies compared 
to the case with $R_0=1.0$~fm. Note, however, that the kinetic energy
by itself is not an observable. 
\begin{table*}
\caption{\label{tab:energies}Kinetic and potential energy contributions
to the GFMC energy at \nxlo{0}, \nxlo{1}, and \nxlo{2} for both cutoffs
and for a particular choice of \nnn{} $D$ and $E$ operators
[Eqs.~\eqref{eq:appvdcoord2} and \eqref{eq:vetdott}].
For $R_0=1.0$~fm, $c_D=0.0$, $c_E=-0.63$, while for $R_0=1.2$~fm,
$c_D=3.5$, $c_E=0.085$~\cite{lynn2016}.
For comparison, we also show results for the Argonne $v_{18}$ \nn{}
interaction supplemented with the UIX \nnn{} interaction.
$V_{\nnn{}}$ stands for the sum of all \nnn{} contributions.
All energies are in MeV.}
{\renewcommand{\arraystretch}{1.30}
\begin{ruledtabular}
\begin{tabular}{lrc;+------}
&&$R_0$ (fm)&\multicolumn{1}{c}{$K$}&\multicolumn{1}{c}{$V_{\nn{}}$}&
\multicolumn{1}{c}{$V_{\nnn{}}$}&\multicolumn{1}{c}{$V_{C,c_1}$}&
\multicolumn{1}{c}{$V_{C,c_3}$}&\multicolumn{1}{c}{$V_{C,c_4}$}&
\multicolumn{1}{c}{$V_{D2}$}&\multicolumn{1}{c}{$V_{E\tau}$}\\
\hline
\multirow{8}{*}{\isotope[3]{H}}&
\multirow{2}{*}{\nxlo{0}}&1.0&60.2(2)&-74.0(2)&&&&&&\\
&&1.2&55.5(1)&-67.8(1)&&&&&&\\
&\multirow{2}{*}{\nxlo{1}}&1.0&46.3(2)&-54.4(2)&&&&&&\\
&&1.2&36.9(2)&-45.0(2)&&&&&&\\
&\multirow{2}{*}{\nxlo{2}}&1.0&42.7(2)&-50.6(2)&-1.32(2)&-0.08(1)&-1.22(2)&-0.53(7)&0.0&0.51(1)\\
&&1.2&37.6(1)&-45.9(1)&-0.87(1)&-0.06(1)&-0.27(1)&-0.35(3)&-0.09(1)&-0.10(1)\\
&&&&&&&&&&\\
&AV18+UIX&&51.4(2)&-59.4(2)&-1.23(1)&&&&&\\
\\
\multirow{8}{*}{\isotope[3]{He}}&
\multirow{2}{*}{\nxlo{0}}&1.0&60.0(1)&-73.0(1)&&&&&&\\
&&1.2&55.0(1)&-67.4(1)&&&&&&\\
&\multirow{2}{*}{\nxlo{1}}&1.0&43.9(3)&-51.5(3)&&&&&&\\
&&1.2&36.4(2)&-44.3(2)&&&&&&\\
&\multirow{2}{*}{\nxlo{2}}&1.0&41.3(3)&-50.5(2)&-1.27(2)&-0.08(1)&-1.16(2)&-0.53(9)&0.0&0.49(1)\\
&&1.2&36.8(1)&-45.1(1)&-0.83(1)&-0.05(1)&-0.26(1)&-0.34(3)&-0.08(1)&-0.09(1)\\
&&&&&&&&&&\\
&AV18+UIX&&50.4(1)&-58.4(1)&-1.20(1)&&&&&\\
\\
\multirow{8}{*}{\isotope[4]{He}}&
\multirow{2}{*}{\nxlo{0}}&1.0&142.0(2)&-193.4(2)&&&&&&\\
&&1.2&132.1(2)&-183.5(2)&&&&&&\\
&\multirow{2}{*}{\nxlo{1}}&1.0&90.2(3)&-115.9(3)&&&&&&\\
&&1.2&73.0(3)&-99.4(2)&&&&&&\\
&\multirow{2}{*}{\nxlo{2}}&1.0&90.9(2)&-116.1(2)&-7.46(4)&-0.41(1)&-6.74(5)&-2.6(2)&0.0&2.34(2)\\
&&1.2&79.9(2)&-106.3(2)&-5.56(4)&-0.30(1)&-1.78(3)&-1.7(2)&-1.24(4)&-0.51(1)\\
&&&&&&&&&&\\
&AV18+UIX&&115.8(1)&-140.4(1)&-6.73(2)&&&&&\\
\end{tabular}
\end{ruledtabular}}
\end{table*}

The trend represented in Fig.~\ref{fig:ekme} is also present in the
radii of the system: See Fig.~\ref{fig:ekmr} and
Table~\ref{tab:ppradii}.
Here we compute the so-called point-proton radii of $A=3,4$ systems:
\begin{equation}
\ev{r_\text{pt}^2}\equiv\Big\langle\Psi_0\Big|\frac{1}{Z}\sum_{i=1}^A
\left(\frac{1+\tau_{z,i}}{2}\right)r_i^2\Big|\Psi_0\Big\rangle\,,
\end{equation}
where $(1+\tau_z)/2$ is a projection operator onto protons and $Z$ is
the number of protons.
However, the measured charge radius includes effects from the charge
densities of the finite-sized nucleons themselves.
The relationship between the point-proton radius and the observable
charge radius $r_\text{c}$ is given by
\begin{equation}
\label{eq:ppradii}
\ev{r_\text{c}^2}=
\ev{r_\text{pt}^2}+\ev{r_p^2}+\frac{N}{Z}\ev{r_n^2}+\frac{3}{4}
\frac{(\hbar c)^2}{m_p^2}\,,
\end{equation}
where $\sqrt{\ev{r_p^2}}=0.8751(61)$~fm is the root-mean-square (rms)
charge radius of the proton~\cite{pdg2016}, $N$ is the number of
neutrons, $\ev{r_n^2}=-0.1161(22)\text{ fm}^2$ is the mean-square
charge radius of the neutron~\cite{pdg2016}, and $m_p$ is the proton
mass.
The last term of Eq.~(\ref{eq:ppradii}) is the so-called Darwin-Foldy
correction to the proton charge radius~\cite{friar1997}.
For larger $A$, there are also spin-orbit corrections to the charge
radius~\cite{ong2010}.
The experimental charge radii are from Ref.~\cite{sick2014}
(\isotope[4]{He} and \isotope[3]{He}) and Ref.~\cite{amroun1994}
(\isotope[3]{H}).

\begin{table}
\caption{\label{tab:ppradii}Point-proton radii as calculated in
Eq.~\eqref{eq:ppradii} at \nxlo{0}, \nxlo{1}, and \nxlo{2} for both
cutoffs for the $A=3,4$ nuclei.
The theoretical uncertainties are from both the GFMC statistical
uncertainties as well as the theoretical uncertainty coming from the
truncation of the chiral expansion as described in
Sec.~\ref{sec:localcheft}.
Experimental values are from
Refs.~\cite{pdg2016,friar1997,sick2014,amroun1994} with
uncertainties calculated using standard propagation of uncertainty
methods.
All radii are in fm.}
{\renewcommand{\arraystretch}{1.20}
\begin{ruledtabular}
\begin{tabular}{l......}
&\multicolumn{2}{c}{\isotope[3]{H}}&\multicolumn{2}{c}{\isotope[3]{He}}&
\multicolumn{2}{c}{\isotope[4]{He}}\\
\hline
$R_0$&
\multicolumn{1}{c}{$1.0\ \text{fm}$}&
\multicolumn{1}{c}{$1.2\ \text{fm}$}&
\multicolumn{1}{c}{$1.0\ \text{fm}$}&
\multicolumn{1}{c}{$1.2\ \text{fm}$}&
\multicolumn{1}{c}{$1.0\ \text{fm}$}&
\multicolumn{1}{c}{$1.2\ \text{fm}$}\\
\hline
\nxlo{0}&1.27(35)&1.27(37)&1.36(56)&1.36(52)&1.02(55)&1.00(53)\\
\nxlo{1}&1.62(10)&1.64(13)&1.92(16)&1.88(18)&1.57(15)&1.53(18)\\
\nxlo{2}&1.55(4)&1.55(6)&1.77(5)&1.77(6)&1.43(5)&1.42(7)\\
Exp&\multicolumn{2}{c}{1.59(10)}&\multicolumn{2}{c}{1.78(2)}&
\multicolumn{2}{c}{1.46(1)}
\end{tabular}
\end{ruledtabular}}
\end{table}

The correlation between the energies and radii of the nuclei are evident
in Fig.~\ref{fig:ekmr} and Table~\ref{tab:ppradii}.
At \nxlo{0}, as the nuclei are significantly overbound, the point-proton
radii are significantly smaller than the values extracted from
experiment.
At \nxlo{1}, with the nuclei underbound, the point-proton radii are too
large.
At \nxlo{2}, with reasonable reproduction of the nuclear binding
energies for the $A=3,4$ systems, the calculated point-proton radii are
in good agreement within both the experimental and theoretical
uncertainties.
Note that the relatively large uncertainty quoted in the point-proton
radius for \isotope[3]{H} extracted from experiment is due to the
relatively large uncertainty in the charge radius for this nucleus:
Compare $r_\text{c}(\isotope[3]{He})=1.973(14)$~fm with
$r_\text{c}(\isotope[3]{H})=1.755(86)$~fm.
The experimental uncertainty is roughly a factor of six larger for
\isotope[3]{H} than for \isotope[3]{He}.

\subsection{More details on distributions}
In addition to energies and radii, we have also calculated one-body
distributions.
The one-body point distributions are defined as
\begin{equation}
\label{eq:pointdens}
\rho_{1,N}(r)\equiv\frac{1}{4\pi
r^2}\Big\langle\Psi_0\Big|\sum_{i=1}^A\frac{1\pm\tau_{z,i}}{2}
\delta(r-|\vb{r}_i-\vb{R}_\text{cm}|)\Big|\Psi_0\Big\rangle\,,
\end{equation}
with $N=p$ (taking the positive sign in the projector
$\tfrac{1+\tau_z}{2}$) giving the point-proton distribution and $N=n$
(taking the negative sign in the projector $\tfrac{1-\tau_z}{2}$) giving
the point-neutron distribution.
When folded with the spatial proton charge distribution, the point
proton distribution is promoted to the charge distribution, which is the
Fourier transform of the charge form factor measured in electron
scattering experiments.
The short-distance behavior of the presented point-nucleon distributions
are not as well constrained, because the high momentum-exchange charge
form factor is challenging to measure and to calculate accurately.
Nevertheless, the charge radius (or point-proton radius) as an
integrated quantity is well constrained by experiment, and our results
reproduce within uncertainties the point-proton radii extracted from
experiment.

\begin{figure}[t]
   \includegraphics[width=\columnwidth]{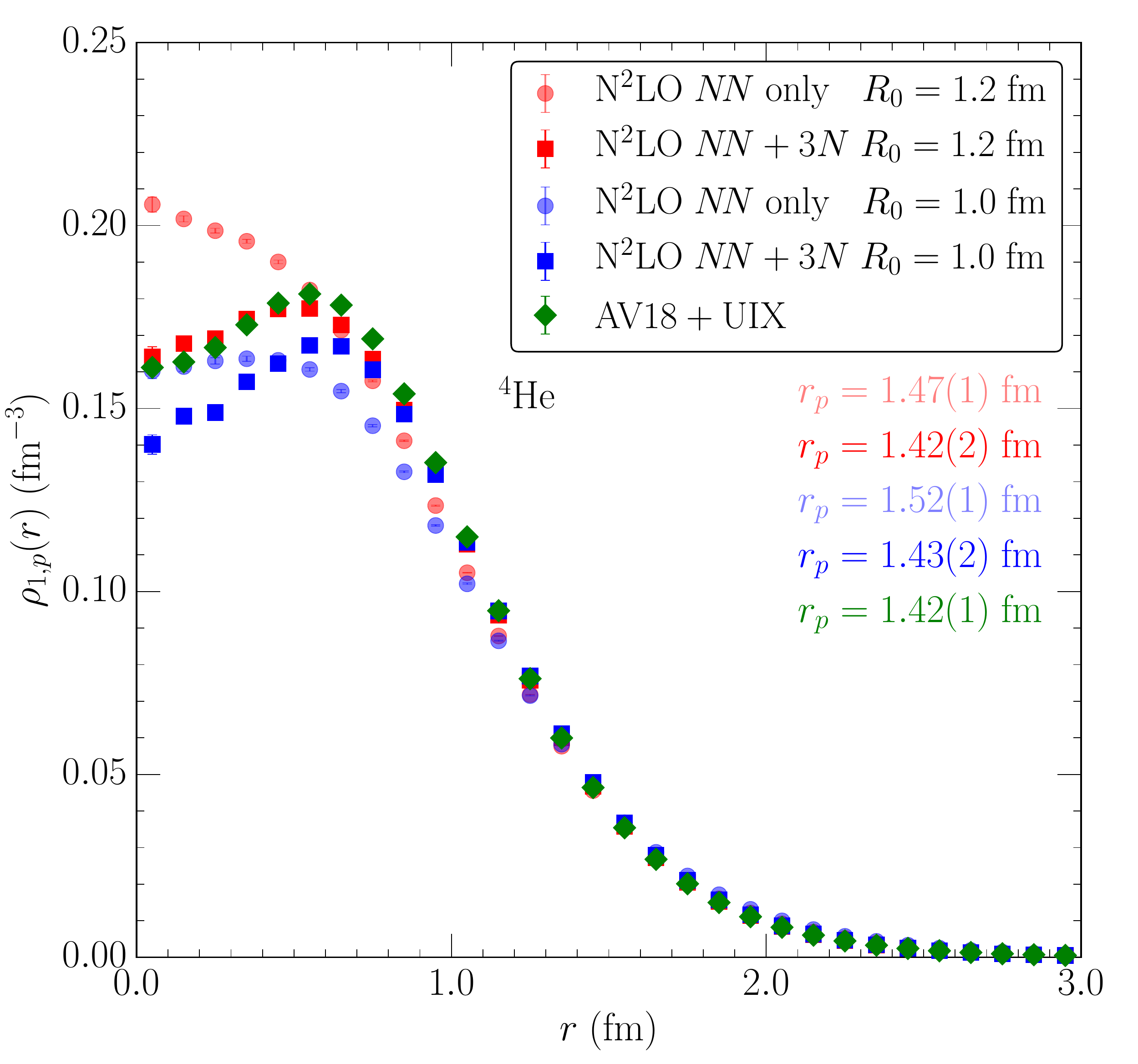}
   \caption{\label{fig:onebd1ptz}
   The one-body proton distribution for \isotope[4]{He} at \nxlo{2} with
   and without \nnn{} interactions for the two different cutoffs we
   consider.
   The darker (lighter) points include (exclude) \nnn{} interactions.
   The corresponding point-proton radii are shown in a color-coded
   fashion to the right.
   The uncertainties quoted for the point-proton radii include only the
   GFMC statistical uncertainties.
   See Table~\ref{tab:ppradii} for more details on the point-proton radii
   including uncertainties from the truncation of the chiral expansion.
   }
\end{figure}

In Fig.~\ref{fig:onebd1ptz}, we show the point-proton distribution in
\isotope[4]{He} for both cutoffs $R_0=1.0,1.2$~fm at \nxlo{2} with and
without the \nnn{} interaction.
The corresponding point-proton radius is shown in a color-coded way on
the right-hand side of the figure.
Though it is not consistent from the EFT point of view to show the
\nxlo{2} results without the \nnn{} interaction, it is nevertheless
instructive to see the effects of the \nnn{} interaction in this way.
One can see that its effect is to increase the density of protons at
intermediate distances from the center of mass ($r\sim1.0$~fm) while
decreasing their density at short distances, yielding a peak at about
$r~\sim0.6$~fm.
The effect of this shift is to bring the overall point-proton radius
into better agreement with the number extracted from the experimental
charge radius.

\begin{figure}[t]
   \includegraphics[width=\columnwidth]{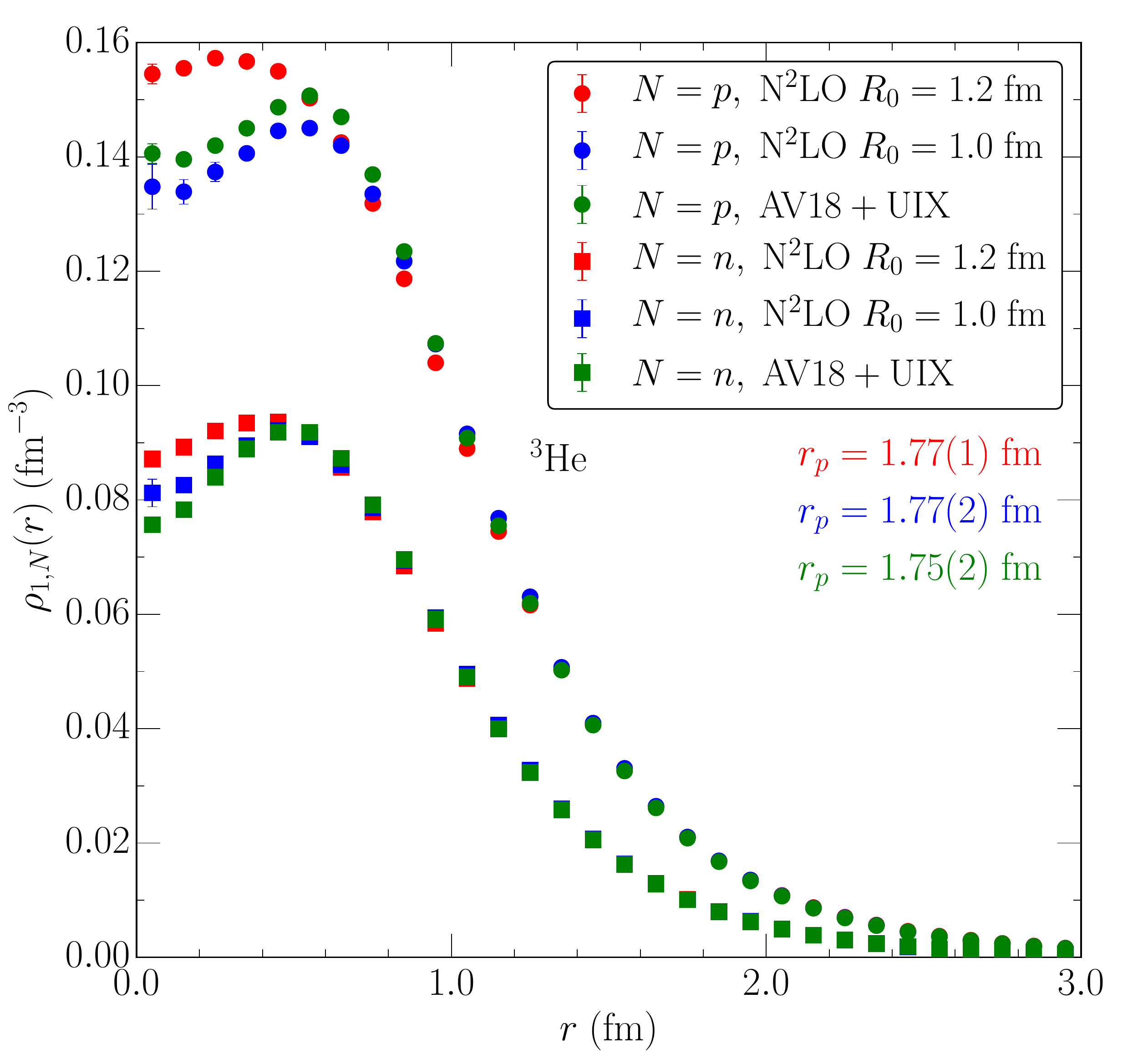}
   \caption{\label{fig:he31bdpndens}
   The one-body proton and neutron distributions for \isotope[3]{He} at
   \nxlo{2} for the two different cutoffs we consider.
   The corresponding point-proton radii are shown in a color-coded
   fashion to the right.
   The uncertainties quoted for the point-proton radii include only the
   GFMC statistical uncertainties.
   See Table~\ref{tab:ppradii} for more details on the point-proton radii
   including uncertainties from the truncation of the chiral expansion.
   }
\end{figure}

In Fig.~\ref{fig:he31bdpndens} we show the one-body point-proton and
neutron distributions for \isotope[3]{He} at \nxlo{2} for both cutoffs.
At short distances from the center of mass, the distributions for the
cutoff $R_0=1.2$~fm demonstrate a softer character: There is a higher
probability of finding either a neutron or a proton at short distances
from the center of mass than is the case for the distributions calculated
with the $R_0=1.0$~fm cutoff.
As is the case for \isotope[4]{He}, only the large-$r$ part of the
distributions can be well constrained, and in this region,
both cutoffs agree.
We also show the corresponding point-proton radii with statistical
GFMC uncertainties only, to demonstrate that integrated quantities such
as the charge radius are essentially cutoff independent for these light
systems at this order of the chiral expansion. Finally, in 
Fig.~\ref{fig:he31bdpndens}, one can see that the proton distribution
is qualitatively twice the neutron distribution, but there are
quantitative differences due to the presence of 
isospin-symmetry-breaking terms in the Hamiltonian.

The point-proton and point-neutron distributions we calculate are
related to the experimentally observable electric charge form factor.
In particular, the longitudinal electric charge form factor is given
by
\begin{equation}
F_L(q)=\frac{1}{Z}\frac{G_E^p(Q_\text{el}^2)\tilde{\rho}_p(q)+
G_E^n(Q_\text{el}^2)\tilde{\rho}_n(q)}{\sqrt{1+Q_\text{el}^2/(4m_N^2)}}\,,
\end{equation}
where $\tilde{\rho}$ are the Fourier transforms of the point-nucleon
distributions defined in Eq.~(\ref{eq:pointdens}), $G_E^{n,p}$ are the
single nucleon electric charge form factors for the neutron $n$ and
proton $p$ and $Q_\text{el}^2$ is the four-momentum squared:
\begin{equation}
Q_\text{el}^2=\vb{q}^2-\omega_\text{el}^2
\end{equation}
with
\begin{equation}
\omega_\text{el}=\sqrt{q^2+m_A^2}-m_A\,.
\end{equation}
Above, $m_N$ and $m_A$ are the average nucleon mass and the mass of the
target nucleus, respectively.
For the single-nucleon charge form factors $G_E^{n,p}$, we use the
parametrizations of Kelly~\cite{kelly2004}, which enforce the correct
asymptotic behavior as $Q^2_\text{el}\to0$ and $Q^2_\text{el}\to\infty$.

\begin{figure}[t]
   \includegraphics[width=\columnwidth]{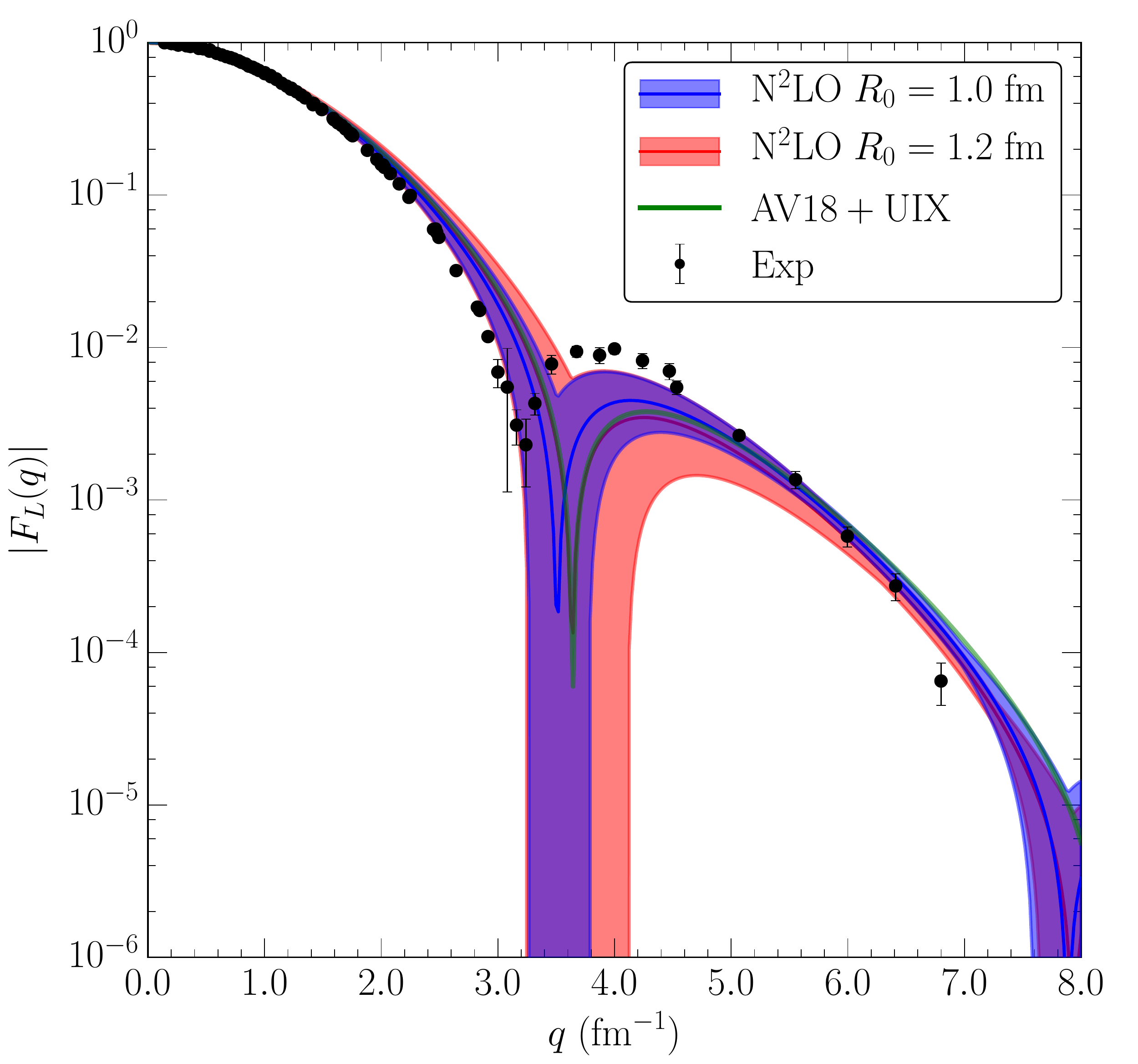}
   \caption{\label{fig:he4cffekm}
   The \isotope[4]{He} longitudinal charge form factor at \nxlo{2} for
   both cutoffs and for the $\text{AV}18+\text{UIX}$ interactions.
   The uncertainty bands include the statistical GFMC uncertainties
   added in quadrature (for the \nxlo{2} results) to the uncertainty
   from the truncation of the chiral expansion as described in the text.
   The data are from an unpublished compilation by Sick based on
   Refs.~\cite{frosch1967,erich1968,mccarthy1977,arnold1978,
   ottermann1985}.
   }
\end{figure}

In Fig.~\ref{fig:he4cffekm}, we present the longitudinal electric charge
form factor for \isotope[4]{He} compared with an unpublished compilation
by Sick~\cite{isickcomp} of the data from
Refs.~\cite{frosch1967,erich1968,mccarthy1977,arnold1978,ottermann1985}.
The figure is log scaled as charge form factors are often plotted,
but this scaling artificially enhances the apparent size of the
uncertainties.
However, the figure should be read as simply that at \nxlo{2} the
uncertainty in the location of the first minimum in the \isotope[4]{He}
charge form factor is roughly 0.6--0.8 fm$^{-1}$.
Note that calculations are performed without two-body currents,
and thus the poorer comparison with data at higher $q$ is somewhat
expected~\cite{carlson2015}.

\section{Summary}
\label{sec:sumandconcl}
In this paper, we presented additional details on and results for
QMC calculations of light nuclei with local chiral \nn{} and \nnn{}
interactions. 
We discussed deuteron properties in detail, employing a soft and a
hard local chiral interaction. We found that local chiral interactions
give a reasonable description of the deuteron binding energy, rms
radius, asymptotic $D/S$ ratio, and quadrupole moment.
Furthermore, local chiral interactions reproduce the experimentally
known first minimum of the deuteron tensor polarization.

We then performed perturbative calculations for both interactions
in the deuteron, using the difference of the \nxlo{2} and \nxlo{1} 
interactions as a perturbation around the \nxlo{1} result. While 
both perturbative series seem to converge to the \nxlo{2} result, 
we found the softer interaction to be more perturbative, as
expected.  

We then presented additional details on our calculations of radii 
and binding energies of the light $A=3,4$ nuclei \isotope[3]{H},
\isotope[3]{He}, and \isotope[4]{He}. For each binding energy
and radius and for both local chiral interactions, we observed an
order-by-order convergence toward the experimental value.
Finally, we discussed proton and neutron distributions for
\isotope[3]{He} and \isotope[4]{He}.

Together with the results of Ref.~\cite{lynn2016}, we have established
QMC methods with local chiral interactions as a versatile tool to
study properties of light nuclei and neutron matter. 

\acknowledgments{
We thank S.~Bacca, P.~Klos, and D.~Lonardoni for useful discussions.
This work was supported by the ERC Grant No. 307986 STRONGINT,
the National Science Foundation Grants No.~PHY-1430152 (JINA-CEE)
and No.~PHY-1404405, the U.S. DOE under Grants No.~DE-AC52-06NA25396
and No.~DE-FG02-00ER41132, the NUCLEI SciDAC program, the Natural 
Sciences and Engineering Research Council (NSERC) of Canada, and the
LANL LDRD program. Computational resources have been provided by 
the J\"ulich Supercomputing Center, the Lichtenberg high performance 
computer of the TU Darmstadt, and Los Alamos Open Supercomputing.
We also used resources provided by NERSC, which is supported by the 
U.S. DOE under Contract No. DE-AC02-05CH11231.
}\\

\appendix*
\onecolumngrid
\section{COMPLETE COORDINATE-SPACE EXPRESSIONS FOR THE
\texorpdfstring{$\boldsymbol{\nnn{}}$}{3N} INTERACTION AT
\texorpdfstring{\nxlo{2}}{N2LO}}
\label{app}
As noted in Sec.~\ref{subsec:nnn}, the TPE parts of the
\nnn{} interaction $V_C$ can be compactly written in terms of the
standard coordinate-space pion propagator $X_{ij}(\vb{r})$ (defined in
that section) and a modified coordinate-space pion propagator
$\mathcal{X}_{ij}(\vb{r})\equiv
X_{ij}(\vb{r})-\frac{4\pi}{m_\pi^2}\drtn{r}\sdots{i}{j}$.
We also define the function $U(r)=1+1/(m_\pi r)$.
Then, the complete TPE part of the interaction can be
written as
\begin{subequations}
\begin{align}
\label{eq:compvcc1}
V_{C,c_1}&=\frac{g_A^2m_\pi^4c_1}{16\pi^2F_\pi^4}
\sum_{i<j<k}\sum_\text{cyc}
U(r_{ij})Y(r_{ij})U(r_{kj})Y(r_{kj})\tdott{i}{k}
\vb*{\sigma}_i\vdot\vb{\hat{r}}_{ij}
\vb*{\sigma}_k\vdot\vb{\hat{r}}_{kj}\,,\\
\label{eq:compvcc3}
V_{C,c_3}&=\frac{g_A^2m_\pi^4c_3}{1152\pi^2F_\pi^4}
\sum_{i<j<k}\sum_\text{cyc}\{\tdott{i}{k},\tdott{k}{j}\}
\{\mathcal{X}_{ik}(\vb{r}_{ik}),\mathcal{X}_{kj}(\vb{r}_{kj})\}\,,\\
\label{eq:compvcc4}
V_{C,c_4}&=-\frac{g_A^2m_\pi^4c_4}{2304\pi^2F_\pi^4}
\sum_{i<j<k}\sum_\text{cyc}[\tdott{i}{k},\tdott{k}{j}]
[\mathcal{X}_{ik}(\vb{r}_{ik}),\mathcal{X}_{kj}(\vb{r}_{kj})]\,.
\end{align}
\end{subequations}
The remaining parts of the interaction are written as
\begin{subequations}
\begin{align}
\label{eq:appvdcoord1}
V_{D1}&=
\frac{g_A c_D m_\pi^2}{96 \pi \Lambda_\chi F_\pi^4}
\sum_{i<j<k}\sum_\text{cyc}\tdott{i}{k}\left[
\vphantom{\frac{8\pi}{m_\pi^2}}
X_{ik}(\vb{r}_{kj})\drtn{r_{ij}}
+X_{ik}(\vb{r}_{ij})\drtn{r_{kj}}-\frac{8\pi}{m_\pi^2}\sdots{i}{k}
\drtn{r_{ij}}\drtn{r_{kj}}\right]\,,\\
\label{eq:appvdcoord2}
V_{D2}&=
\frac{g_A c_D m_\pi^2}{96 \pi \Lambda_\chi F_\pi^4}
\sum_{i<j<k}\sum_\text{cyc}\tdott{i}{k}\left[
\vphantom{\frac{4\pi}{m_\pi^2}}X_{ik}(\vb{r}_{ik})
-\frac{4\pi}{m_\pi^2}\sdots{i}{k}\drtn{r_{ik}}\right]
\Bigl[\drtn{r_{ij}}+\drtn{r_{kj}}\Bigr]\,,
\end{align}
\end{subequations}
\begin{subequations}
\begin{align}
\label{eq:vetdott}
V_{E\tau}&=\frac{c_E}{\Lambda_\chi F_\pi^4}\sum_{i<j<k}\sum_\text{cyc}
\tdott{i}{k}\drtn{r_{kj}}\drtn{r_{ij}}\,,\\
\label{eq:veoneone}
V_{E\mathbbm{1}}&=\frac{c_E}{\Lambda_\chi
F_\pi^4}\sum_{i<j<k}\sum_\text{cyc}
\drtn{r_{kj}}
\drtn{r_{ij}}\,,\\
\label{eq:veproj}
V_{E\mathcal{P}}&=\frac{c_E}{\Lambda_\chi
F_\pi^4}\sum_{i<j<k}\sum_\text{cyc}\mathcal{P}\,
\drtn{r_{kj}}
\drtn{r_{ij}}\,.
\end{align}
\end{subequations}
We remind the reader that the projection operator $\mathcal{P}$ is
defined in Eq.~\eqref{eq:projector}.
We note that some differences exist between these expressions compared
with those in Ref.~\cite{tews2016}.
Under the change $\sum_{\pi(ijk)}
\to\sum_\text{cyc}$, Eqs.~(\ref{eq:compvcc1}) and (\ref{eq:vetdott}) to
(\ref{eq:veproj})
pick up an additional factor of 2 and Eq.~(\ref{eq:appvdcoord1}) picks
up an additional term with $i\leftrightarrow k$.
In addition, Eqs.~(\ref{eq:compvcc3}) and (\ref{eq:compvcc4}) pick up
factors of $\tfrac{1}{2}$ and $\tfrac{1}{2\ii{}}$, respectively, from
the replacements
$\tdott{i}{j}=\tfrac{1}{2}\{\tdott{i}{k},\tdott{k}{j}\}$ and
$\vb*{\tau}_i\vdot(\vb*{\tau}_j\cross\vb*{\tau}_k)
=\tfrac{1}{2\ii{}}[\tdott{i}{k},\tdott{k}{j}]$.
\twocolumngrid
\bibliography{longqmcwxeft_final}
\end{document}